\documentclass[useAMS,usenatbib]{mn2e}
\usepackage{graphicx} 
\usepackage{amssymb}
\usepackage{color}
\newcommand{\gasoline}{{\small\rm GASOLINE }}
\newcommand {\hi} {{\rm H}\,{\small\rm I}}
\newcommand {\hicap} {{\rm H}\,{\scriptsize\rm I}}

\newcommand {\ovi} {{\rm O}\,{\small\rm VI}}

\newcommand {\siiii} {{\rm Si}\,{\small\rm III}}
\newcommand {\siiv} {{\rm Si}\,{\small\rm IV}}
\newcommand {\cii} {{\rm C}\,{\small\rm II}}
\newcommand {\ciii} {{\rm C}\,{\small\rm III}}
\newcommand {\civ} {{\rm C}\,{\small\rm IV}}
\newcommand {\kms} {\,{\rm km\,s}^{-1}}

\newcommand {\pc} {\,{\rm pc}}

\newcommand {\kpc} {\,{\rm kpc}}

\newcommand {\cmmq}{\,{\rm cm^{-2}}}

\newcommand {\kmskpc} {\,{\rm km\,s}^{-1}\,{\rm \kpc}^{-1}}

\newcommand {\de}{^{\circ}}

\newcommand {\msun}{\,{\rm M}_\odot}

\newcommand {\msunsqp}{\,{\rm M}_\odot \, {\rm pc}^{-2}}
\newcommand{\vsun}{\,{v}_\odot}

\newcommand{\yr}{\,{\rm yr}}

\newcommand{\Gyr}{\,{\rm Gyr}}
\newcommand{\K}{\,{\rm K}}

\newcommand {\msunyr}{\,{{\rm M}_\odot\,\rm yr}^{-1}}

\newcommand {\vlsr}{v_{\rm LSR}}

\newcommand{\avg}[1]{\left< #1 \right>} % for average

\title[The effect of feedback on a Milky Way-like galaxy] 
{The effect of stellar feedback on a Milky Way-like galaxy and its gaseous halo}
\author[Marasco et al.]
{
	Antonino Marasco$^{1}$\thanks{E-mail:marasco@astro.rug.nl},
	Victor P. Debattista$^{2, 3}$,
	Filippo Fraternali$^{1, 4}$,\newauthor
	Thijs van der Hulst$^{1}$,
        James Wadsley$^5$,
        Thomas Quinn$^6$,
        Rok Ro{\v s}kar$^7$
	\\
	$^{1}$Kapteyn Astronomical Institute,  Postbus 800, 9700 AV, Groningen, The Netherlands\\
	$^{2}$Jeremiah Horrocks Institute, University of Central Lancashire, Preston PR1 2HE, UK\\
	$^{3}$Department of Physics, University of Malta, Tal-Qroqq Street, Msida MSD 2080, Malta\\
	$^{4}$Department of Physics \& Astronomy, University of Bologna, via Berti Pichat 6/b, 40127, Bologna, Italy\\
	$^{5}$Department of Physics and Astronomy, McMaster University, Hamilton, Ontario L8S 4M1, Canada\\
	$^{6}$Astronomy Department, University of Washington, Box 351580, Seattle, WA 98195-1580, USA\\
	 $^{7}$Research Informatics, Scientific IT Services, ETH Z\"urich,
  Weinbergstrasse 11, 8092, Z\"urich, Switzerland
}
\begin{document}

\date{Accepted 2015 May 28.  Received 2015 May 12; in original form 2015 February 27}

\pagerange{\pageref{firstpage}--\pageref{lastpage}} \pubyear{2014}
\maketitle
\label{firstpage}

\begin{abstract}
  We present the study of a set of N-body+SPH simulations of a
  Milky Way-like system produced by the radiative cooling of hot gas
  embedded in a dark matter halo. The galaxy and its gaseous halo
  evolve for 10 Gyr in isolation, which allows us to study how
  internal processes affect the evolution of the system.  We show how
  the morphology, the kinematics and the evolution of the galaxy are
  affected by the input supernova feedback energy $E_{\rm SN}$, and we
  compare its properties with those of the Milky Way.  Different
  values of $E_{\rm SN}$ do not significantly affect the star
  formation history of the system, but the disc of cold gas gets
  thicker and more turbulent as feedback increases.  Our main result
  is that, for the highest value of $E_{\rm SN}$ considered, the
  galaxy shows a prominent layer of extra-planar cold
  ($\log(T/\K)\!<\!4.3$) gas extended up to a few kpc above the disc
  at column densities of $10^{19}\cmmq$.  The kinematics of this
  material is in agreement with that inferred for the \hi\ halos of
  our Galaxy and NGC 891, although its mass is lower.  Also, the
  location, the kinematics and the typical column densities of the hot
  ($5.3\!<\!\log(T/\K)\!<\!5.7$) gas are in good agreement with those
  determined from the \ovi\ absorption systems in the halo of the Milky
  Way and external galaxies.  In contrast with the observations,
  however, gas at $\log(T/\K)\!<\!5.3$ is lacking in the
  circumgalactic region of our systems.
\end{abstract}


\section{Introduction}\label{introduction}

Numerical simulations constitute a fundamental tool to understand the
processes of galaxy formation and evolution.  Simulations of isolated
systems have proved to be very powerful to investigate the physics on
galactic scales, and they have been widely used in the last decade to
interpret the secular evolution of discs \citep[e.g.][]{Debattista+06,
  Combes08}, to study the role of stellar feedback
\citep[e.g.][]{Stinson+06, DallaVecchiaSchaye08,
  DallaVecchiaSchaye12}, and to understand the effect of major and
minor mergers on the morphology and dynamics of galaxies
\citep[e.g.][]{VillalobosHelmi11, Bois+11}.

Simulations can be used to address a long-standing question in the
theory of galaxy evolution, namely how disc galaxies are able to
sustain star formation for a Hubble time without consuming their gas
reservoir.  In our Galaxy, for instance, the star formation rate (SFR)
of the solar neighborhood has remained almost constant for the last
$\sim10\Gyr$ \citep{Twarog80,AumerBinney09}, implying that the gas
consumed by star formation is continuously replenished.  More
generally, there is evidence that galaxies, at all times, must have
accreted gas at a rate proportional to their SFR density
\citep{Hopkins+08,FraternaliTomassetti12}.  However, there is little observational evidence
for cold (\hi) gas accretion occurring onto galaxies at the required rate
\citep[e.g.][]{Sancisi+08}.

$\Lambda$CDM cosmological simulations made a major breakthrough in
this field by revealing the modes by which galaxies may accrete their gas
from the cosmic web.  Low-mass halos are expected to accrete material through the
so-called `cold-mode', where filaments of gas at relatively low
temperature ($\lesssim10^5\K$) are able to penetrate the halos down to
the centre and feed directly the formation of the central galaxy
\citep[e.g.][]{Keres+09}.  In massive halos, instead, accreting
filaments get shock-heated and settle into a rarefied atmosphere of
plasma at the virial temperature of the halo, often called
\emph{coronae} \citep{WhiteRees78, Crain+10}.  Because of the
hierarchical formation of structures, the high-redshift universe is
expected to be dominated by the cold-mode accretion, which is
progressively replaced by the `hot-mode' as structures grow
\citep{Keres+05, DekelBirnboim06}.  Massive disc galaxies at redshift
$z=0$ - including the Milky Way - are therefore expected to be
surrounded by a large amount of rarefied hot gas - the so-called `coronae' - which would
constitute a vast reservoir of material to fuel star formation.
 
Simulations also reveal that feedback from stars and active galactic
nuclei (AGN) plays a fundamental role in the evolution of galaxies.
In most cases, stellar feedback on a galactic scale slows down the
star formation in the disc (`negative' feedback) \citep[][hereafter S06]{Stinson+06}
and produces outflows of hot gas from the disc to the halo
\citep{DallaVecchiaSchaye08}.  On a cosmological scale, stellar and
AGN feedback are used to solve the mismatch between the halo mass
function and the galaxy mass function by quenching star formation in
both low-mass and high-mass systems.  Additionally, stellar feedback
pollutes the circumgalactic medium (CGM\footnote{In this paper the terms \emph{CGM} and \emph{corona} have two different meanings: the first refers to all the gas that surrounds a galaxy disc, the second specifically refers to the hot ($\sim$ virial-temperature) plasma that settles in equilibrium around the galaxy and is built by the hot-mode accretion.}) with metals
and produces absorption-line systems consistent with those observed
around galaxies in the spectra of background sources \citep[e.g.][]{Stinson+12}.  In general, the
tuning of feedback in state-of-the-art cosmological simulations
nowadays allows a much better match with the observations with respect
to the recent past \citep[see][]{Vogelsberger+14, Schaye+15}.

Despite the importance of feedback processes, there is no general
consensus on how they must be implemented in numerical simulations.
This is due to two reasons.  On the one hand, these processes occur on
scales (both spatial and temporal) that are commonly unresolved by the
current generation of simulations.  This leads to a plethora of
different numerical recipes that attempt to approximate the physics of
feedback on such `sub-grid' scales, often delivering different
outcomes despite the similarities of the initial conditions
\citep{Scannapieco+12}.  On the other hand, the amount of mechanical
and thermal energy deposited into the surrounding gas by these
processes is very difficult to constrain observationally.  This
implies that the feedback energy is treated as a free parameter, which
can be tuned ad-hoc to reproduce the observations. Even though
state-of-the art feedback recipes are attempting to address these
issues \citep[e.g.][]{Keller+14}, a final solution has still to be
found.

In this paper we investigate how the properties of a Milky Way-like
system surrounded by its CGM are affected by stellar feedback.  Our
system is produced by the radiative cooling of a rotating hot gas
component (a corona) embedded in a NFW dark matter halo.  The gas at
the bottom of the potential well cools and settles into a disc,
eventually reaching the density required to form stars and producing
feedback from supernovae and winds.  The system constituted by the
cold, star-forming disc and the hot CGM evolves in isolation for
$10\Gyr$.  Thus the properties of the final object are affected only
by the interplay between these two components.  This allows us to
study the impact of internal processes (like stellar feedback from the
disc and radiative cooling of the corona) on the evolution of a
Milky Way-like galaxy.  The novelties of our work with respect to
previous studies of simulations of isolated systems are the
followings: a) our galaxies are surrounded by an extended hot corona,
which mimics the hot-mode epoch of gas accretion that galaxies like
the Milky Way have experienced since redshift of $\sim2$ (i.e.
lookback time of $\sim10\Gyr$); and b) we present a direct comparison
of the gas component with real data of the Milky Way by treating the
simulated system like an observation.

The paper is structured as follows.  In Section \ref{simulations} we
present the simulations used in this work.  In Section \ref{results}
we describe the morphology, the mass distribution, the kinematics and
the gas circulation in these systems.  In Section \ref{comparison} we
focus on the gaseous components and mimic observations of the
simulations in order make a direct comparison between them and the
Milky Way.  In Section \ref{discussion} we discuss the results
obtained.  We present our conclusions in Section \ref{conclusions}.

\section{The simulations}
\label{simulations}

The simulations are run with the N-body + SPH code \gasoline
\citep{Wadsley+04}.  The initial conditions are set as in
\citet{Roskar+08}.  We considered a spherical NFW dark matter halo
with virial radius ($r_{200}$) of $200\kpc$ and virial mass of
$10^{12} \msun$.  Note that these values are perfectly compatible with
recent estimates of the Galactic dark matter halo
\citep[e.g.][]{McMillan11}.  The halo has an embedded spherical corona
of hot gas containing $10\%$ of the total mass and following the same
density distribution.  This corona is initially in hydrostatic
equilibrium for an adiabatic equation of state, and has a spin
parameter of $\lambda = 0.065$ \citep{bul_etal_angmom_01,
  mac_etal_07b} with specific angular momentum $\propto R$, where $R$
is the cylindrical radius.  Both gas and dark matter components are
described by $1\times10^6$ particles.  Gas particles have initial mass
of $1.4\times10^5\msun$ and softening length of $50\pc$, while dark
matter particles have softening of $\sim100\pc$ and masses of either
$10^6\msun$ or $3.5\times10^6 \msun$, depending on whether they are
inside or outside $r_{200}$.  The initial gas is comprised of a
mixture of hydrogen and helium, with no metals.

At the beginning of the simulation radiative cooling is
  switched-on, allowing the central, densest region of the hot gas to
  cool and settle into a disc.  Star formation and stellar feedback
  from SNII, SNIa and stellar winds are implemented according to the
  recipes of S06, and the system is followed up to
  $t=10\Gyr$.  We use a base timestep $\Delta t = 10$ Myr with a
  refinement parameter $\eta = 0.175$ such that timesteps satisfy
  $\delta t = \Delta t/2^n < \eta\sqrt{\epsilon a_g}$, where
  $\epsilon$ is the softening parameter (set to 50 pc for baryons and
  100 pc for dark matter) and $a_g$ is the acceleration a particle
  experiences.  Gas particles also satisfy the timestep condition
  $\delta t_{gas} = \eta_{courant} h/[(1+\alpha)c + \beta\mu_{max}]$,
  where $\eta_{courant} = 0.4$, $h$ is the SPH smoothing length,
  $\alpha$ is the shear coefficient, which is set to 1, $\beta=2$ is
  the viscosity coefficient and $\mu_{max}$ is described in
  \citet{Wadsley+04}.  The SPH smoothing kernel encloses the 32 nearest
  neighbours.  We use an opening angle for gravity of $\theta = 0.7$.
  Star formation is triggered at a number density $n > 0.1 {\mathrm
    cm}^{-3}$, temperature $T < 15,000$ K and provided the gas
  particle is part of a converging flow; efficiency of star formation
  is 0.05 per dynamical time.  Star particles form with an initial
  mass of $1/3$ that of the gas particle.  Once the mass of a gas
  particle drops below $1/5$ of its initial mass, the remaining mass
  is distributed amongst the nearest neighbors.  Star particles are
  assumed to constitute an entire stellar population with a
  Miller-Scalo \citep{MillerScalo79} initial mass function.  Feedback
  from SNII follows a blast-wave recipe where thermal energy is
  injected into the surrounding gas and, in addition, gas particles
  within the blast-wave radius have their cooling switched off for a
  time of the order of $\sim10^7\yr$ (for details see S06).  Stellar feedback also pollutes the interstellar medium with metals, whose main impact in the simulation is to affect
  the cooling timescale of the gas.
  We adopt a metallicity-dependent cooling function using the
  prescription of \citet{Shen+10a}.  Metal cooling rates
  depend on density, temperature and metallicity and computed from
  tabulated rates.  Metallicity cooling would lead to very low
  temperatures ($T \sim 10\K$) if allowed to proceed unhindered.
  However below $\sim 1000\K$ the Jeans mass then becomes comparable
  to $\msun$ for reasonable densities, which is much below our mass
  resolution.  In order to prevent the artificial fragmentation that
  would result, we follow \citet{Agertz+09a} in setting a pressure
  floor $P = 3 G \epsilon^2 \rho^2$.

As we are interested in distinguishing between gas particles that have
been ejected into the halo by stellar feedback and those that remain
in the CGM throughout, we do not allow metals (or thermal energy) to
diffuse from a given gas particle to the surrounding ones.  With this
choice, all particles have zero metal abundance as long as they are
not directly affected by SN feedback and stellar winds from the disc.
In order to test whether this choice severely affects our results, we
re-ran one of the simulations (F40, see Section \ref{snapshots}) by
implementing metal and thermal diffusion and we found no significant
differences in terms of star formation history and star/gas surface
density profile with respect to the non-diffusive version.  Therefore
we preferred to neglect thermal and metal diffusion in the
simulations.

The version of \gasoline used in this work does not include the most
recent improvements made by \citet{Hopkins13} or \citet{Hobbs+13} in
the SPH numerical scheme.  This choice makes it easier to compare our
findings with the results obtained in previous studies.  Hopkins
pointed out that, in astrophysical contexts, sub-grid physics is what
primarily shapes the outcome of a simulation rather than the numerical
scheme adopted.  Our simulated galaxies do not seem to show
overcooling in their circumgalactic medium, an issue that seems to be
caused by the inefficient phase mixing of the classical SPH scheme.
In addition, the lack of significant differences between the run with
and without thermal diffusion indicates that the details of gas mixing
are not crucial to our setup.

\subsection{Star formation history and masses}
\label{snapshots}

Using the setup described, we produce four different models: F80, F40,
F10 and F2.5.  Each model uses the same initial conditions but differs
by the amount of (thermal) energy injected by SNe II into the
interstellar medium (ISM), with F80 being the most energetic model
($80\%$ of the canonical $10^{51}$ erg per supernova is transferred to
the ISM).  The main physical parameters of the four galaxies after
$10\Gyr$ of evolution are presented in Table \ref{snapshots}.  Note
that, in terms of stellar and halo masses, all our simulated galaxies
are comparable to the Milky Way \citep[e.g.][]{Sofue+11}.

\begin{figure}
\begin{center}
\includegraphics[width=0.5\textwidth]{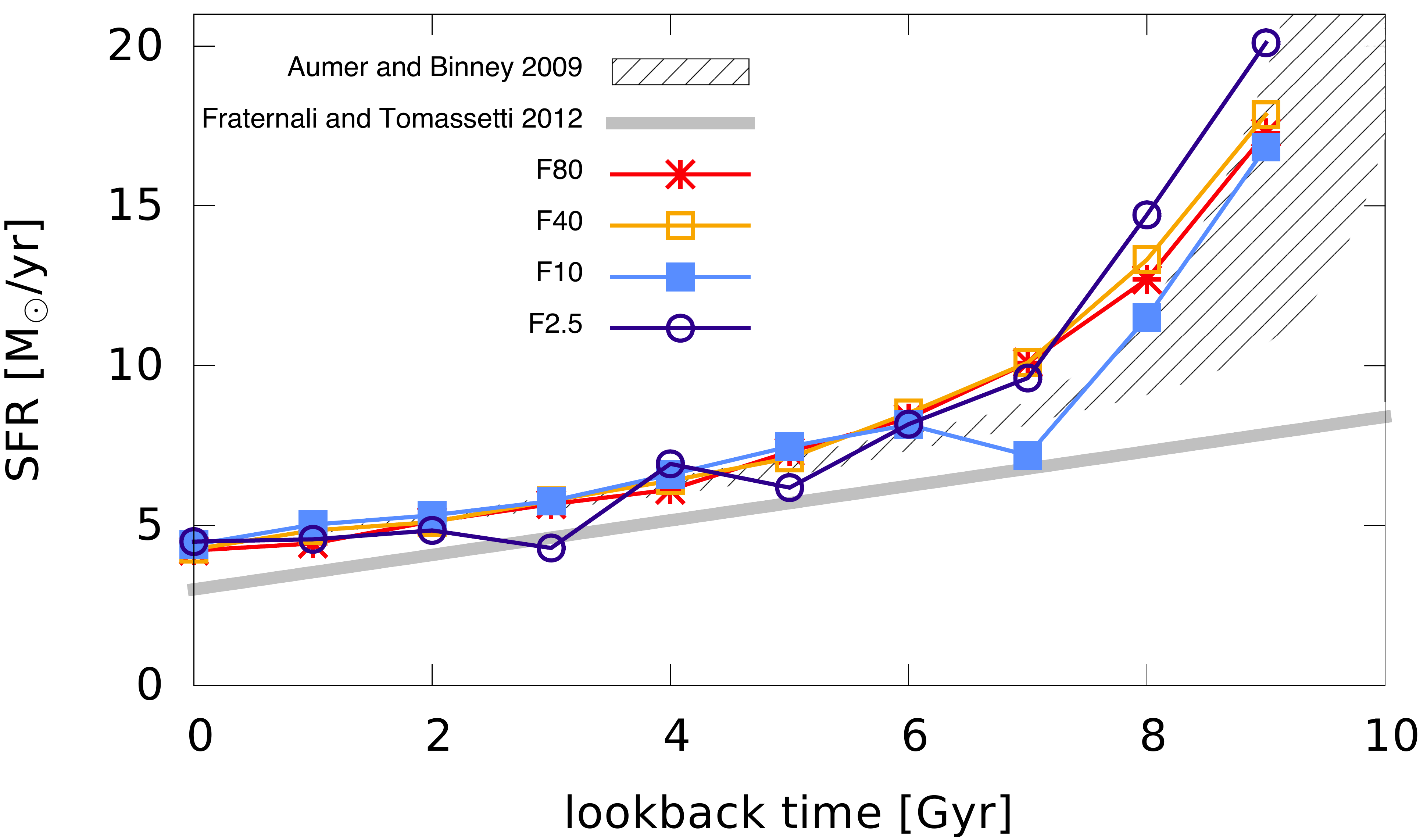}
\caption{Comparison between the star formation history of the
  simulated galaxies (thin lines with points) and that of the Milky
  Way determined by \citet{AumerBinney09} (shaded region) and
  \citet{FraternaliTomassetti12} (thick grey line).}
\label{SFH}
\end{center}
\end{figure}

Fig.\,\ref{SFH} compares the star formation history (SFH) of all the
simulations with those derived for the Milky Way by
\citet{AumerBinney09} (double exponential model) and
\citet{FraternaliTomassetti12}.  The former SFH is based on the
kinematics and colours of the stars in the solar neighborhood, while
the latter is valid for the whole Galaxy and is obtained by assuming a linear trend with lookback time. 
Note that the model of Aumer \& Binney is normalised to the
SFR of our systems at t=10 Gyr, while \citet{FraternaliTomassetti12}
assume a current SFR of $3\msunyr$.  In all models the SFR is in good
agreement with that inferred by \citet{AumerBinney09} at all times,
and only slightly above that predicted by Fraternali \& Tomassetti for
the last $\sim7\Gyr$.  This is an important starting point for the
detailed comparison between the Galaxy and the simulations.

\begin{table*}
\centering
\caption[]{Properties of the simulated galaxies after $10\Gyr$ of evolution.} 
\label{snapshots}
\begin{tabular}{lccccl}
\hline
\hline
{\bf Simulation}          & \emph{F80}	 &\emph{F40}		&\emph{F10}	 &\emph{F2.5} & units \\
\hline
\hline
$M_{\rm{HI+H_2}}$		&$2.9\times10^9$	&$2.5\times10^9$	 &$2.4\times10^9$	 &$2.5\times10^9$	& $\msun$ \\
$M_{\rm stars}$	&$5.8\times10^{10}$&$5.9\times10^{10}$&$6.1\times10^{10}$&$6.2\times10^{10}$&$\msun$ \\
SF rate$^a$ 		&$4.2$			&$4.2$			 &$4.4$			 &$4.5$			& $\msunyr$ \\
%$E_{\rm SN}$			&$0.80$			&$0.40$			 &$0.10$			 &$0.025$			& $10^{51}$ erg\\ 			
\hline
\end{tabular}
\begin{footnotesize}
\begin{list}{}{}
\item[$^a$]  evaluated by considering the amount of stars born in the last 50 Myr. 
\end{list}
\end{footnotesize}
\end{table*}

\subsection{Neutral gas fraction}
\label{neutralfraction}

Since the simulations did not use radiative transfer, there is no
direct way to distinguish between neutral and ionised gas.  In order
to estimate the neutral and ionised gas fractions, we assume simple
collision ionisation equilibrium (CIE), and define as `neutral' the
gas particles with a temperature below $2\times10^4\K$
\citep[see][]{SutherlandDopita93}.  The neutral hydrogen mass is
then derived by assuming a universal hydrogen abundance fraction of
$0.75$.  We make no distinction between atomic and molecular gas in
the simulations.

A more elaborate approach is the following.  Using $N$-body+SPH
cosmological simulations post-processed with radiative transfer,
\citet{Rahmati+13} found that, at each epoch, a relation between
particle density and photo-ionization rate exists.  We use this
relation (tuned for redshift $z=0$) to derive the neutral gas fraction
for each gas particle in the simulations, assuming an UV background
radiation field based on the model of \citet{HaardtMadau01} \citep[for
details see Appendix A of][]{Rahmati+13}.  The comparison between the
total maps of neutral gas derived with these two approaches revealed
almost no differences, thus for simplicity we chose to adopt the
former simpler method.

\section{Results}
\label{results}

\subsection{Morphology}
\label{morphology}

\begin{figure*}
\begin{center}
\includegraphics[width=\textwidth]{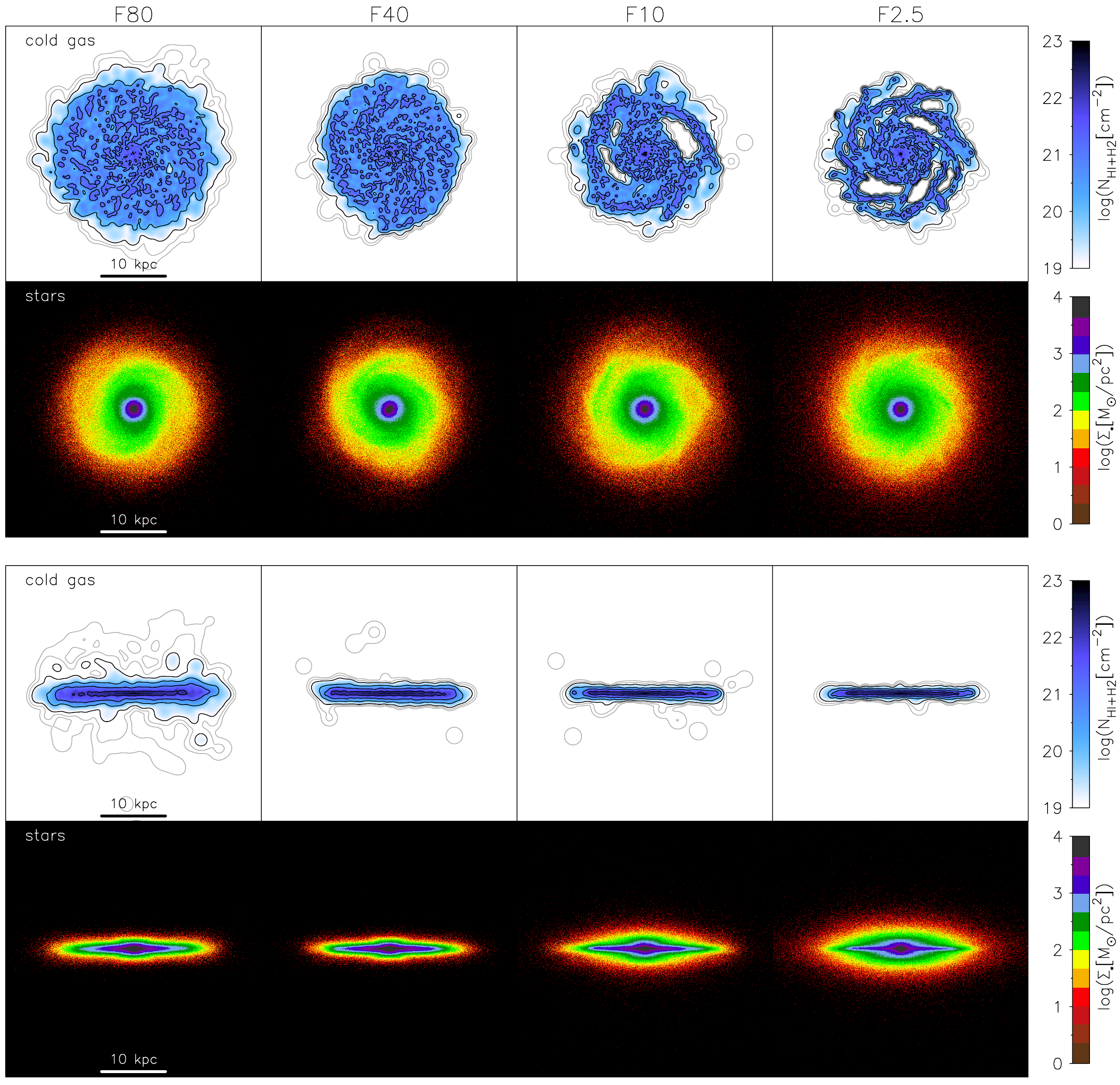}
\caption{Final images of cold gas and stars of the four simulated
  galaxies after 10 Gyr of evolution. All maps are on the same scale.
  \emph{First row:} total maps of cold gas as seen in a face-on
  projection. Black contours are at column densities of $10^{19}$,
  $10^{20}$, $10^{21}$, $10^{22} \cmmq$; grey outermost contours are
  at $10^{18}$ and $10^{18.5}\cmmq$. \emph{Second row:} maps of the
  stellar component as seen in a face-on projection. \emph{Third and
    fourth rows}: as the first and second rows, but the galaxies are
  seen in a edge-on projection.}
\label{maps}
\end{center}
\end{figure*}

Fig.\,\ref{maps} shows the final images of the neutral hydrogen
(atomic+molecular) and of the stellar component, in face-on and
edge-on projections, for all the simulated galaxies analyzed.  The
maps of the gas component are derived by smoothing each particle using
a Gaussian kernel with FWHM equal to the smoothing length of that
particle.  This gives to these maps a `smooth' appearance that is not
present in the maps of the stellar component.

The face-on view reveals that the gaseous discs are extended roughly
as much as their respective stellar component, ranging from $12\kpc$
in F2.5 up to $15\kpc$ in F80 at the column density level of
$10^{19}\cmmq$ (outermost black contour).  Sizes do not increase
further at $10^{18}\cmmq$ (outermost grey contour), suggesting that
the cold gas discs are truncated and gas at larger radii is too hot to
be neutral.  This truncation is not due to the temperature threshold
adopted to define the neutral gas phase, as maps derived by using a
more refined approach (see Section \ref{neutralfraction}) show the
same features at these column densities.  Spirals are visible in all
the gas maps, but the contrast between arm and inter-arm regions
increases with decreasing feedback.  An extreme case is F2.5, where
the inter-arm regions show extended holes in the neutral gas
distributions.  When seen face-on, stellar discs are relatively more
similar to each other.  Spiral features in the gaseous and stellar
component of F2.5 are well correlated.

Compared to the size of the \hi\ disc of the Milky Way \citep[$30\kpc$
at the column density of $10^{19}\cmmq$, see][]{KalberlaKerp09}, the
discs of cold gas in the simulations are too small.  In general, they
appear to fall outside the \hi\ mass-scale relation
\citep[e.g.][]{BroeilsRhee97}. Since in the models the discs of cold
gas are produced by the cooling of the spinning hot halos, their size
will depend on how the angular momentum is spatially distributed in
these media.  We have assumed that rotation in the corona is
distributed as $\sim R$, which is arbitrary and probably influences
the size of the gas disc.  Additionally, mergers and cold flows may
also modify the angular momentum distribution in the corona, and in
fact \citet{Wang+14} successfully reproduced the \hi\ mass-scale
relation by using cosmological simulations in $\Lambda$CDM paradigm.
Hence, we do not consider the size of the \hi\ discs as a prediction
of the simulations.

The most striking features arise when the systems are projected to an
edge-on view.  Gaseous discs respond to the change in feedback by
modifying their thickness.  This happens because high values of
$E_{\rm SN}$ imply a larger injection of thermal energy inside the
disc, which produces more turbulence in the ISM and consequently
thickens the gas layer.  A few extra-planar gas features are visible
already in F40 and F10 at column density lower than $10^{19}\cmmq$;
however, in F80 a thick extra-planar component is visible above
$10^{19}\cmmq$ and is even more remarkable at $10^{18}\cmmq$, where it
extends up to $\sim10\kpc$ from the midplane.  This thick extra-planar
component is remarkably similar to that observed in \hi\ for the
nearby disc galaxy NGC\,891 \citep{Oosterloo+07}, although its column
density is about one order of magnitude lower.  In Section
\ref{extra-planar} we show that not only the morphology but also the
kinematics are similar.  The stellar components follow the opposite
trend, as stellar discs get significantly thicker (i.e., hotter) as
feedback decreases.  
In Section \ref{feedbackimpact} we show that our runs with lower feedback tend to form more stars at early times. These stars have a longer time to heat vertically, ending up at larger heights.
A more detailed analysis of this feature will be presented elsewhere.

\subsection{Mass distribution and kinematics}
\label{massdistandkin}

\begin{figure*}
\begin{center}
\includegraphics[width=\textwidth]{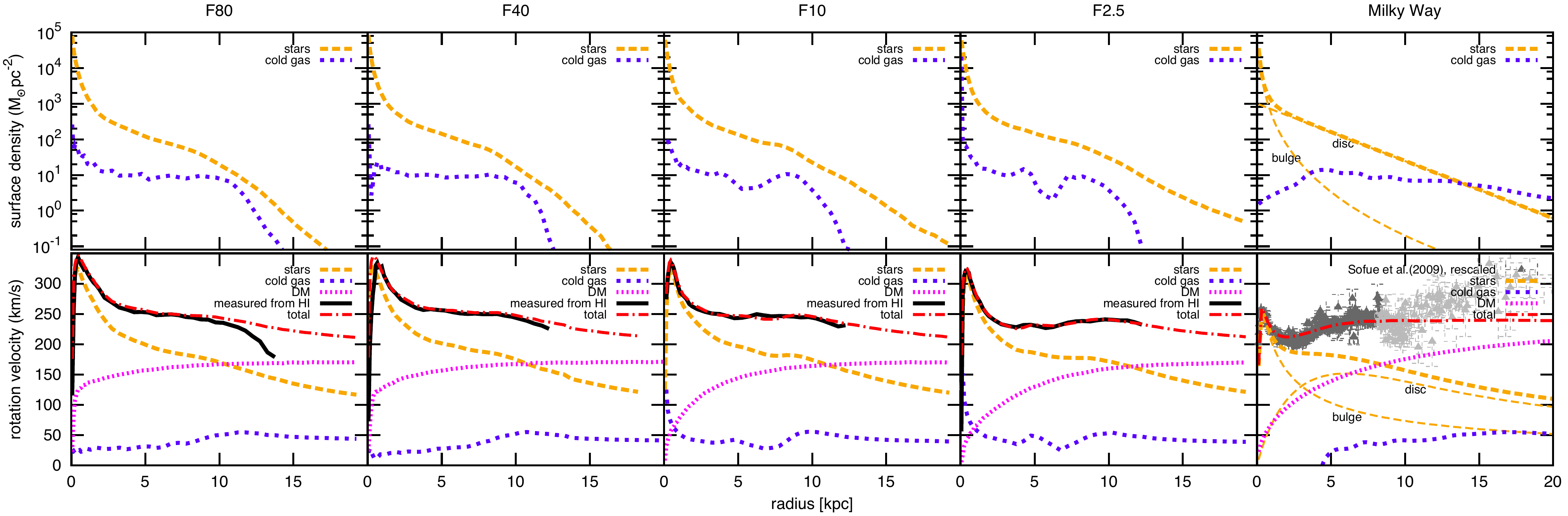}
\caption{\emph{First row:} surface density profiles of the various
  components for the simulated galaxies (first four columns), compared
  with that of the Milky Way (last column). \emph{Second row}: \hicap\
  rotation curve and contribution of the various components to the
  circular velocity for the same galaxies.  The Milky Way's rotation
  curve is taken from \citet{Sofue+09} but it has been rescaled for
  the following values of the Galactic constants: $\vsun\!=\!239\kms$,
  $R_{\odot}\!=\!8.3\kpc$ \citep{McMillan11}.  The cold gas
  (\hicap+H$_2$, corrected for He) surface density of the Milky Way is
  taken from \citet{bm98}, while the stellar surface density and the
  circular velocities are derived from our mass decomposition (see
  text for details).}
\label{massdecomp}
\end{center}
\end{figure*}

The mass distribution and the kinematics of the simulated galaxies are
derived by using an approach similar to the `tilted ring' method that
is often adopted to model the velocity fields of \hi\ observations
\citep[e.g.][]{Begeman+87}.  We focus on the disc of neutral gas, and
divide it into a series of concentric rings, each one with a given
distance with respect to the centre.  This latter is unique and it is
given by the mass centre of the dark matter distribution.  A generic
ring at distance $R$ from the centre is oriented perpendicular to the
angular momentum vector of the neutral gas particles located at that
distance from the centre.  Therefore, the difference of our approach
with respect to the classical tilted-ring method is that the rings
follow the actual three dimensional distribution of gas particles
rather than the inclination and position angle inferred from the
velocity field.  We follow the orientation of each ring in order to
properly evaluate the rotation velocity, the velocity dispersion and
the mass surface density for all the components as a function of
radius.  This approach can easily capture symmetric warps of the gas
component, although non-axisymmetric features may not be treated
properly.  However, none of the models show warps in the neutral gas
distribution (see Fig.\,\ref{maps}).  Also, since this approach relies
on the neutral gas alone to infer the inclination of the various
rings, it can fail if the disc of stars and gas are significantly
misaligned.  This is also not the case for these models.

Fig.\,\ref{massdecomp} shows the gas and stellar surface densities
(top row) and the decomposed rotation curves of cold gas (bottom row),
and a comparison with a mass-decomposition of our Galaxy.  To derive
this latter, we assumed that the Galactic gravitational potential is
produced by four components: a stellar bulge, an double-exponential
stellar disc, a NFW dark matter halo and a disc of cold gas.  Despite
the strong evidence for a `peanut-shaped' boxy-bulge in the Milky Way
\citep[e.g.][]{Dwek+95, Nataf+10, MwilliamZoccali10, Ness+12}, a de
Vaucouleurs bulge provides a very good fit to the Galaxy's rotation
curve \citep{Sofue+09}.  As modelling the stellar distribution and
kinematics of the inner Galaxy is beyond the purpose of this work, for
simplicity we decided to adopt the classical de Vaucouleurs profile to
describe the Galactic bulge.  We fitted the parameters of these
components to the Milky Way's rotation velocity data of
\citet{Sofue+09}, which we rescaled to more up-to-date values of the
Galactic constants $\vsun\!=\!239\kms$ and $R_{\odot}\!=\!8.3\kpc$
\citep{McMillan11}.  As the rotation curve in the outer disc is very
uncertain, we assumed that the rotation velocity flattens at
$R\!>\!R_{\odot}$ and discarded the data points in that region
(light-grey points in Fig.\,\ref{massdecomp}).  We used the cold gas
(\hi+H$_2$, plus a correction for the He) surface density of
\citet{bm98} and we kept it fixed in our fit.  In addition, we
constrained the total baryonic surface density at $R\!=\!R_{\odot}$ to
be in the range $54.2\pm4.9\msunsqp$ \citep{Read14}.  The details of
this mass decomposition will be presented in Marasco \& Fraternali (in
prep).

The cold gas (\hi+H$_2$) surface density profiles of the simulated
systems are all flat at the level of $\sim10\msunsqp$ (or
$\sim10^{21}\cmmq$), and increase by one order of magnitude close to
the centre where star-forming gas piles up.  All profiles are
truncated, as at about $R\!=\!10\kpc$ the surface density decreases by
two orders of magnitude in a small (2-3 kpc) region.  This is
different from what has been found in the Milky Way, where the cold
gas fades gently as the radius increases.  All simulated stellar
profiles show the presence of a mass concentration in the centre due
to a bulge.  We notice that the central stellar surface density
increases slightly with feedback, from $6\times10^4\msunsqp$ in F2.5
to $8\times10^4\msunsqp$ in F80.  Outside the region of the bulge,
stars distribute in a disc that seems to show, in all simulated
systems, a double exponential profile, with the break between the two
slopes occurring approximately where gaseous discs are truncated.
This is not surprising: stars located beyond the gaseous discs must
have been transported to that location by radial migration via
transient spiral corotation capture \citep{SellwoodBinney02,
  Roskar+08a}. A break in the slope of stellar profile has been
observed in several galaxies \citep[e.g.][]{Kregel+02} and it is
possibly present in the Milky Way as well \citep{Sale+10, Minniti+11}. Note that the Galactic \hi\ disc also seems to be
truncated at $R\simeq15\kpc$ \citep{KalberlaDedes08}, although this
truncation is shallower than that shown by our simulated galaxies.
Truncated \hi\ discs are sometimes observed in external galaxies, but they are thought to be caused by photo-ionization from the extragalactic UV background \citep[e.g.][]{Maloney+93} rather then by an effective drop in cold gas density.

The rotation curves, derived by measuring the actual rotation velocity
of cold gas particles in the various rings (black lines), show several
interesting features.  All rotation profiles are steeply rising up to
$350-400\kms$ and then decline and flatten to about $250\kms$ beyond
$R\!\simeq\!3\kpc$.  The Milky Way rotation curve has a similar shape,
but the innermost peak reaches a lower value (slightly above
$250\kms$) and it flattens to a slightly lower rotational speed, whose
precise value depends on the choice of the local standard of rest
rotational velocity \citep[here assumed to be $239\kms$,
see][]{McMillan11}.  Using the particle distribution in the simulation
it is possible to compute the gravitational potential and therefore to
derive the circular rotation for all the mass components separately.
In all models, the dynamics of the system is driven by the stellar
component out to the radius where the gaseous disc is truncated.
Beyond that point, the dark matter becomes the dominant component,
while gaseous discs are not dynamically dominant at any radius as is
expected in this type of galaxies.  As shown in the lower panels of
Fig.\,\ref{massdecomp}, the neutral gas is in perfect circular
rotation except in the outermost regions of F80.  However, as we show
is Section \ref{extra-planar}, part of the cold gas in these regions is
extra-planar and rotates at a slower speed than the gas in the
midplane.

The shape of the dark matter profile deserves mention.  While the
outermost part is similar in all models, the innermost profile rises
significantly more steeply when higher feedback energy is coupled to
the ISM. This corresponds to a higher mass concentration, baryonic and
non-baryonic, in the centre of galaxies with high feedback, and to a
higher peak in rotational velocity.  This result is surprising as it
is opposite to what we expected.  Higher feedback has been
historically invoked to wash out the mass built-up in the centre of
systems \citep[the so-called \emph{core-cusp} problem,
e.g.][]{deBlok10}: as baryonic matter is removed from the system
centre by stellar feedback, dark matter follows it
\citep[e.g.][]{Navarro+96a, ReadGilmore05, OgiyaMori11}.  Our
simulations show the opposite trend, as the central stellar surface
densities increases with $E_{\rm SN}$ and consequently compressing the
dark matter more in the systems with larger feedback output.  
This indicates that the feedback scheme adopted is not very efficient in removing low-angular momentum baryons from the system centres: in fact, it promotes their growth.
We discuss the impact of stellar feedback on our simulations in more detail in Section \ref{feedbackimpact}.

\begin{figure}
\begin{center}
\includegraphics[width=0.4\textwidth]{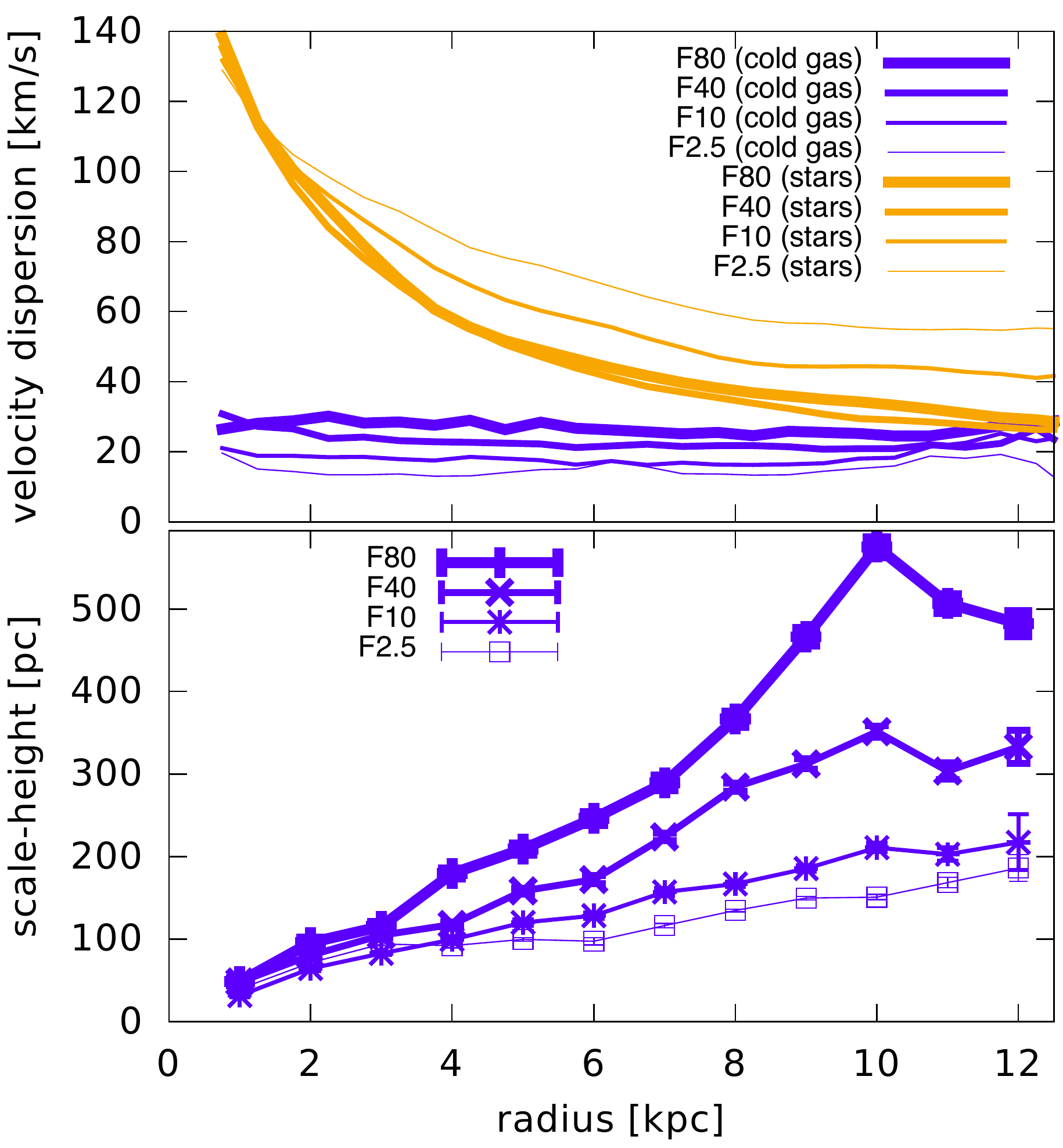}
\caption{\emph{Top panel:} velocity dispersion profiles for both stars
  and cold gas in all the models. \emph{Bottom panel:} scale-height of
  the cold gas as a function of radius in the simulations.}
\label{vdispscale}
\end{center}
\end{figure}

The top panel of Fig.\,\ref{vdispscale} shows the stellar and neutral
gas velocity dispersion profile, $\sigma_*$ and $\sigma_{\rm gas}$,
for the different models, computed as
\begin{eqnarray}
\sigma_*(R) & = & \sqrt{\frac{1}{3}\left(\sigma^2_R(R) + \sigma^2_z(R) + \sigma^2_\phi(R)\right)}\\
\sigma_{\rm gas}(R) & = & \sqrt{\frac{1}{3}\left(\sigma^2_R(R) + \sigma^2_z(R) + \sigma^2_\phi(R)\right)+\frac{K_{\rm B}T}{m_{\rm H}}}
\end{eqnarray}
where $\sigma_R$, $\sigma_z$ and $\sigma_\phi$ are
particle-to-particle velocity dispersions in the various directions
(radial, vertical and azimuthal), averaged over all neutral gas
particles in each ring, $T$ is the mass-averaged temperature of these
particles, and $m_{\rm H}$ is the hydrogen mass.

In all models, $\sigma_{\rm gas}$ is flat and its value depends on the
feedback used, with F2.5 being the least turbulent system.  F80 has
very turbulent motions so that $\sigma_{\rm gas}$ increases up to
$25-30\kms$, $\sim2-3\times$ larger than that measured for the \hi\ in
our Galaxy and in nearby disc galaxies with similar star formation
rates \citep[e.g.][]{Boomsma+08}.  The velocity dispersion of the
stellar component is larger than that of the gas and it increases in
the inner regions where the bulge dominates the total surface density.
As the edge-on maps suggest (bottom panels of Fig.\,\ref{maps}), the
thickening of the stellar discs in the models with lower feedback
corresponds to a general increase in $\sigma_*$.  At $R\!=\!15\kpc$,
there is a factor of $\sim2.5$ of difference in $\sigma_*$ between
F2.5 and F80.

The bottom panel of Fig.\,\ref{vdispscale} shows the scale-height of
cold gas as a function of $R$ for the various models.  All simulated
galaxies show a significant flaring, which is more prominent in the
models with higher feedback.  The \hi\ scale-height in the inner disc
of the Milky Way is between 100 and 200 pc \citep{DickeyLockman90}, so
the cold gaseous disc of F80 is thicker than that of our Galaxy.
However, we point out that the gas velocity dispersion, and therefore
the scale-height, are resolution dependent.  These two quantities are
expected to correlate for discs in hydrostatic equilibrium, and in SPH
simulations gravity - and therefore hydrostatic equilibrium - cannot
be described accurately on scales of the order of the gravitational
softening $\epsilon$.  In all the simulations we adopted
$\epsilon\!=\!50\pc$ for the stars and gas, implying that gravity
deviates from Newtonian on scales comparable to the scale-heights of
\hi\ discs in galaxies.  Thus we stress that, even though the
trends shown by the scale-height and velocity dispersion with feedback
in the models are certainly genuine, the actual values of these
quantities are probably overpredicted as the gravitational restoring
force to the plane is underpredicted within a few $\epsilon$ from the
midplane.
  
\subsection{Time-evolution of stars and neutral gas}
\label{timevol}

\begin{figure*}
\begin{center}
\includegraphics[width=\textwidth]{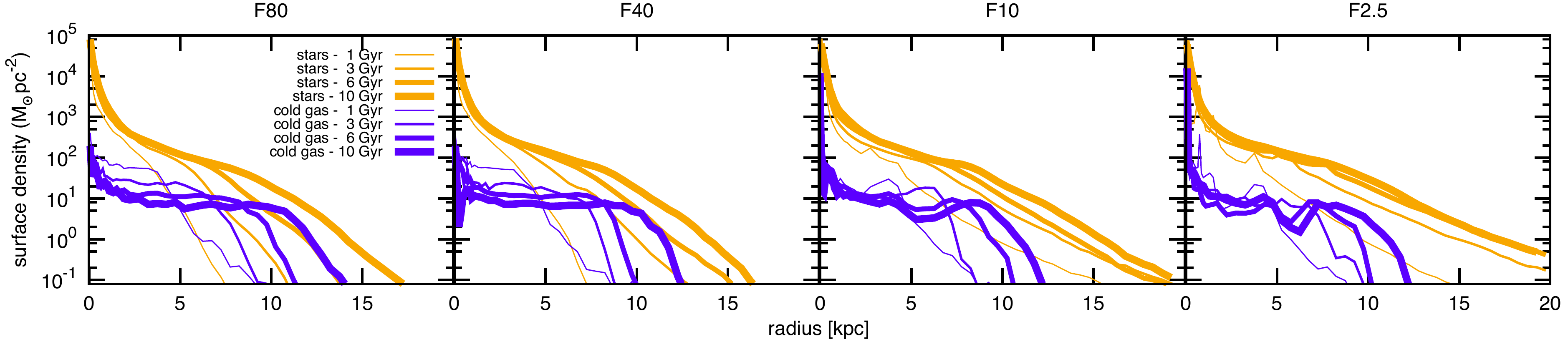}
  \caption{Time-evolution of stellar surface density and cold gas surface density in the simulations. Thicker lines
    represent later stages of evolution, as indicated in the leftmost
    panel.}
\label{evolution}
\end{center}
\end{figure*}

We evolve the models for $10\Gyr$.  Fig.\,\ref{evolution} shows the
evolution of the systems.  Thinner lines represent earlier stages of
the evolution (i.e., larger lookback times).  In all systems both
gaseous and stellar discs form inside-out.  Also, at all times,
stellar discs are systematically more extended than cold gaseous
discs, with the exception of F80 and F40 at $t\!=\!1\Gyr$.  Stars are
born within the cold gas disc then radial migration drives stars
outwards, producing the systematic mismatch in sizes.  We already
noticed in Section \ref{massdistandkin} that stellar discs show a
double exponential profile, with the break in the slope occurring at
the point where gaseous discs are truncated.  Fig.\,\ref{evolution}
shows that this feature occurs at any time.  Stellar migration seems
to continuously transfer stars from the place where they are born, the
gaseous discs, to outer radii, thus producing the observed double
exponential distribution in the stellar profile
\citep[see][]{Roskar+08}.  Along with stellar discs, gaseous discs
also grow inside-out, extending to larger radii at later times.  This
happens in the simulations because the gas cooling time increases
outwards where the density is lower, thus cold gas at large radii
becomes available only during the later stage of the evolution of the
system \citep{Roskar+08}.

\section{Direct comparison with the gas of the Milky Way}
\label{comparison}

In this Section, we focus on the gas component.  Our approach is to
simulate an `observation' by considering an observer placed inside the
disc of the simulated galaxies at $R\!=\!8.5\kpc$ from the centre, and
studying how - from this specific point of view - gas particles
distribute as a function of the angular position in the simulated sky
and of the line-of-sight velocity in the local standard of rest (LSR).
This `pseudo' LSR is determined by averaging the motion of about $100$
star particles around the selected point of view.  We stress that this
approach allows a direct comparison between the gas component of the
simulated systems and that of our Galaxy, for which only 3D
position-velocity information are available.

We compare this simulated observation with two different datasets.
The neutral gas ($\log(T)\!<\!4.3$) is compared with the all-sky \hi\
emission data of the LAB Survey \citep{Kalberla+05}, whereas gas at
higher temperatures ($4.3\!<\!\log(T)\!<\!5.7$) is compared with the
absorption line measurements of ionized species of \citet{Lehner+12},
\citet{Sembach+03} and \citet{Savage+03}.

\subsection{The \hi\ disc} 
\label{HIdisc}

The LAB Survey is an all-sky datacube of Galactic ($|v_{\rm
  LSR}|\!<\!400\kms$) \hi\ emission.  In order to compare the
simulations with the LAB data, we produce a simulated datacube (a
`modelcube') by using the following procedure:
\begin{itemize}
\item we evaluate the pseudo-LSR as described above;
\item in this reference frame, we evaluate the longitude $l$, the
  latitude $b$ and the line-of-sight velocity $v_{\rm LSR}$ of each
  particle. The latitude $b\!=\!0\de$ is chosen to be aligned to the
  disc midplane, so that this reference frame effectively mimics the
  Galactic coordinate system.
\item We place each particle in a 3D grid ($l,b,v$) with pixel size
  equal to that of the LAB data;
\item we smooth the neutral gas content of each particle in velocity
  according to its thermal velocity dispersion $(k_{\rm
    B}T/\mu/m_p)^\frac{1}{2}$.
\item we smooth the neutral gas content of each particle according to
  the desired spatial resolution (see below);
\item we convert each pixel of this modelcube to a brightness
  temperature by assuming an optically thin regime.
\end{itemize}
The spatial smoothing process deserves further discussion.  The
`natural' resolution of the simulations is the SPH kernel size, that
is the distance between a given particle and its $32^\mathrm{nd}$
neighbour.  If we smooth each particle separately to a spatial
resolution equal its SPH kernel size, we assure that the whole
simulated box is sampled without discontinuities.  However, the
smoothing lengths vary significantly from the galaxy centers, where
particles crowd and the resolution is high, to the outskirt of the
systems, where particles are rarer and the resolution is significantly
lower.  This effect is dramatically amplified by the inner projection
used: we found that, on the midplane, the angular resolution varies
from less than $2\de$ at $l\!=\!0\de$ (i.e. in the direction of the
galactic centre) up to $\sim15\de$ at $l\!=\!180\de$.  In addition,
the angular resolution gets significantly worse at higher latitudes.
As we are comparing our modelcubes with data that have a fixed angular
resolution, we decided not to use this smoothing method and instead to
smooth each particle to a fixed angular resolution by using a 2D
Gaussian kernel with a FWHM of $8\de\times8\de$.

\begin{figure*}
\begin{center}
\includegraphics[width=\textwidth]{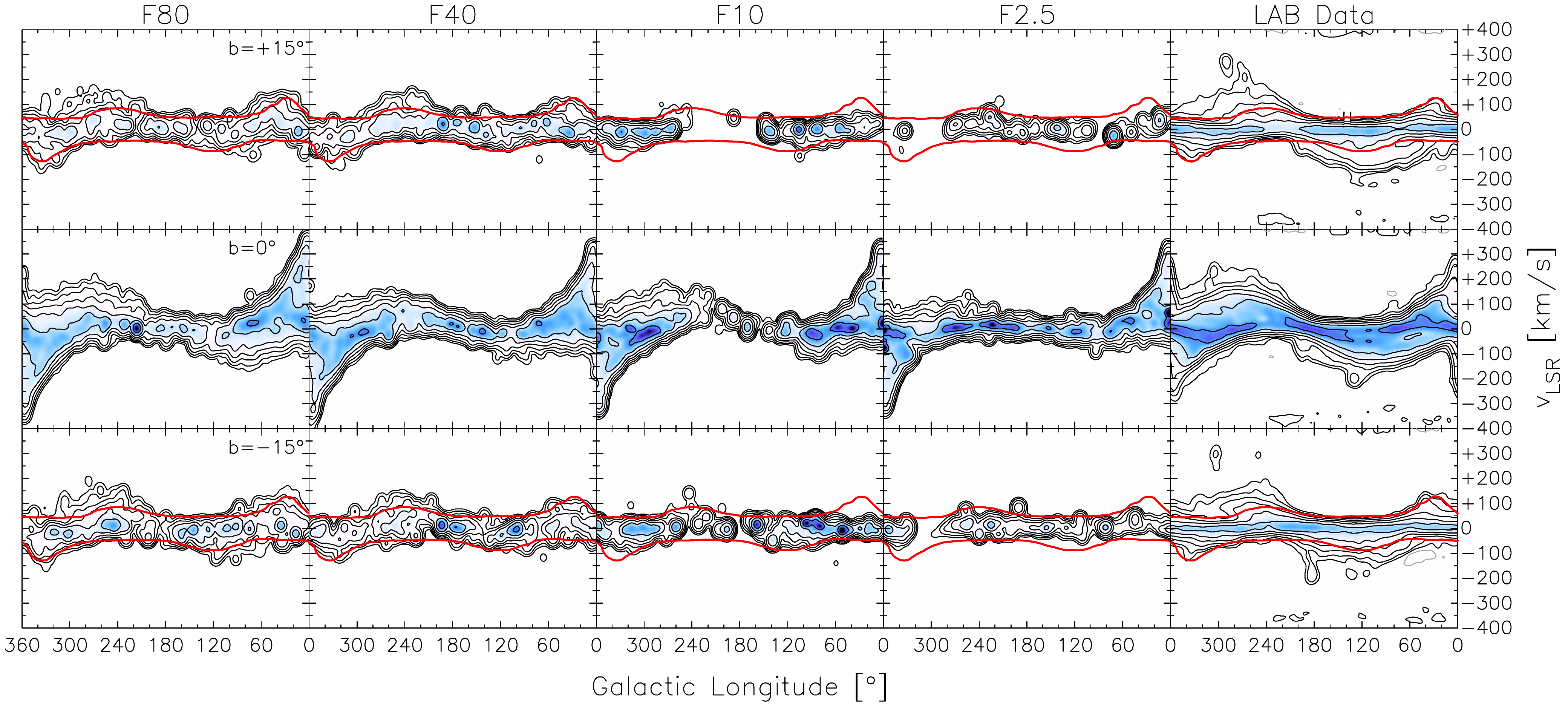}
\caption{Longitude-velocity slices taken at three different latitudes
  (as indicated on top of the leftmost panels) for cold gas in the
  simulated galaxies (columns 1 to 4) and the LAB data of the Milky
  Way (5th column). Black contours are at brightness temperatures
  ranging from $0.01\K$ to $40.96\K$ in multiples of 4.  In the LAB
  data an additional contour at $-0.01\K$ is shown in grey. The red
  contour represents the \hi\ emission at the level of $0.01\K$ from a
  model of Galactic disc with a scale-height of 150 pc.  All panels
  have been smoothed to $8\de$ of angular resolution and $10\kms$ of
  velocity resolution.}
\label{lv_HI}
\end{center}
\end{figure*}

Fig.\,\ref{lv_HI} shows longitude-velocity ($l\!-\!v$) slices taken at
three different latitudes ($\pm15\de$ and $0\de$) for our modelcubes,
and compare them with the LAB data smoothed at the same spatial
resolution ($8\de$). The \hi\ line profiles of all cubes have been
smoothed in velocity by using a Gaussian kernel with a FWHM of $10\kms$ in order to emphasize
the low-level emission in the LAB data. The midplane ($b\!=\!0\de$)
panels show that the velocity spread of the \hi\ emission decreases
with decreasing feedback, which can be explained by the drop in
$\sigma_{\rm gas}$ discussed in Section \ref{massdistandkin}.  The
main differences between the simulated galaxies and the Milky Way at
this latitude are that \hi\ emission in the former reaches larger
velocities around $l\!=\!0\de$ and lower velocities around
$l\!=\!90\de,~ 270\de$.  This happens respectively because a) in the
simulations the rotation curves peak at higher velocities than those
reached in our Galaxy; b) the gaseous discs are smaller than that of
the Milky Way.  These considerations apart, from the slice at
$b\!=\!0\de$ alone it is not possible to decide which model better
resembles the Galactic \hi\ emission.

Moving to different latitudes (top and bottom rows of
Fig.\,\ref{lv_HI}) it becomes clear that, as feedback decreases, the
$l\!-\!v$ plots lose more of the sinusoidal shape that is visible in
the data.  This shape is produced by the rotation of the \hi\ layers
located immediately above the midplane and results from the vertical
extent and the kinematics of this gas.  In the Milky Way, most of the
low-level emission visible at these latitudes is due to the
extra-planar \hi.  This medium, which is thought to be produced by the
Galactic fountain \citep{MFB12}, rotates slower than the gas in the
disc and extends up to a few kpc above the midplane
\citep{MarascoFraternali11}.  Both these features significantly
enhance the sinusoidal shape visible in the LAB data.  As a
comparison, on each panel we overlaid a thick red contour representing
the \hi\ emission (at the level of $0.01\K$) predicted for a
differentially rotating disc extended up to $25\kpc$ and with a
scale-height of 150 pc, approximately the value estimated for the thin
\hi\ disc of our Galaxy out to $R\!=\!R_{\odot}$
\citep{DickeyLockman90}.  For simplicity, we do not attempt to model
the warp in the outer \hi\ Galactic disc \citep[e.g.][]{Levine+06},
which at any rate would affect only a small portion of our modelcube
($40\de\!<\!l\!<\!160\de$, $-200\!<\!\vlsr\!<\!-50\kms$).  Clearly,
most of the low-level emission of the LAB data is located outside this
contour.  Neither F2.5 nor F10 have extra-planar gas (Section
\ref{morphology}) and their $l\!-\!v$ emission is confined within the
contour of the thin disc model.  F40 also does not have extra-planar
gas, but the average thickness of its disc is larger than 150 pc
(bottom panel of Fig.\,\ref{vdispscale}) thus some emission leaks
outside the red contour.  Finally, F80 has a layer of extra-planar gas.
Even though at this stage we do not know what the kinematics of this
layer is, the fact that $l\!-\!v$ emission of F80 is the one that
resembles the LAB data best would suggest that the extra-planar
material rotates slower than the gas in the thin disc.  We show that
this is the case in the next Section.

\subsection{The extra-planar \hi}
\label{extra-planar}

In the last 10-15 years, deep \hi\ observations reaching column
densities of $\sim10^{19}\cmmq$ have proved that star-forming disc
galaxies often show a vertically extended extra-planar \hi\ component
\citep[or `thick \hi\ discs', e.g.][]{Fraternali+02,
  Barbieri+05,Gentile+13}.  Our Galaxy is not an exception
\citep{MarascoFraternali11}.  A detailed modelling of this component
reveals that its rotational velocity decreases with increasing
distance from the midplane.  These kinematics can be explained by a
model of the galactic fountain \citep{Bregman80} interacting with an
ambient medium at a lower specific angular momentum than that of the
disc \citep{FB08}.

\begin{figure}
\begin{center}
\includegraphics[width=0.4\textwidth]{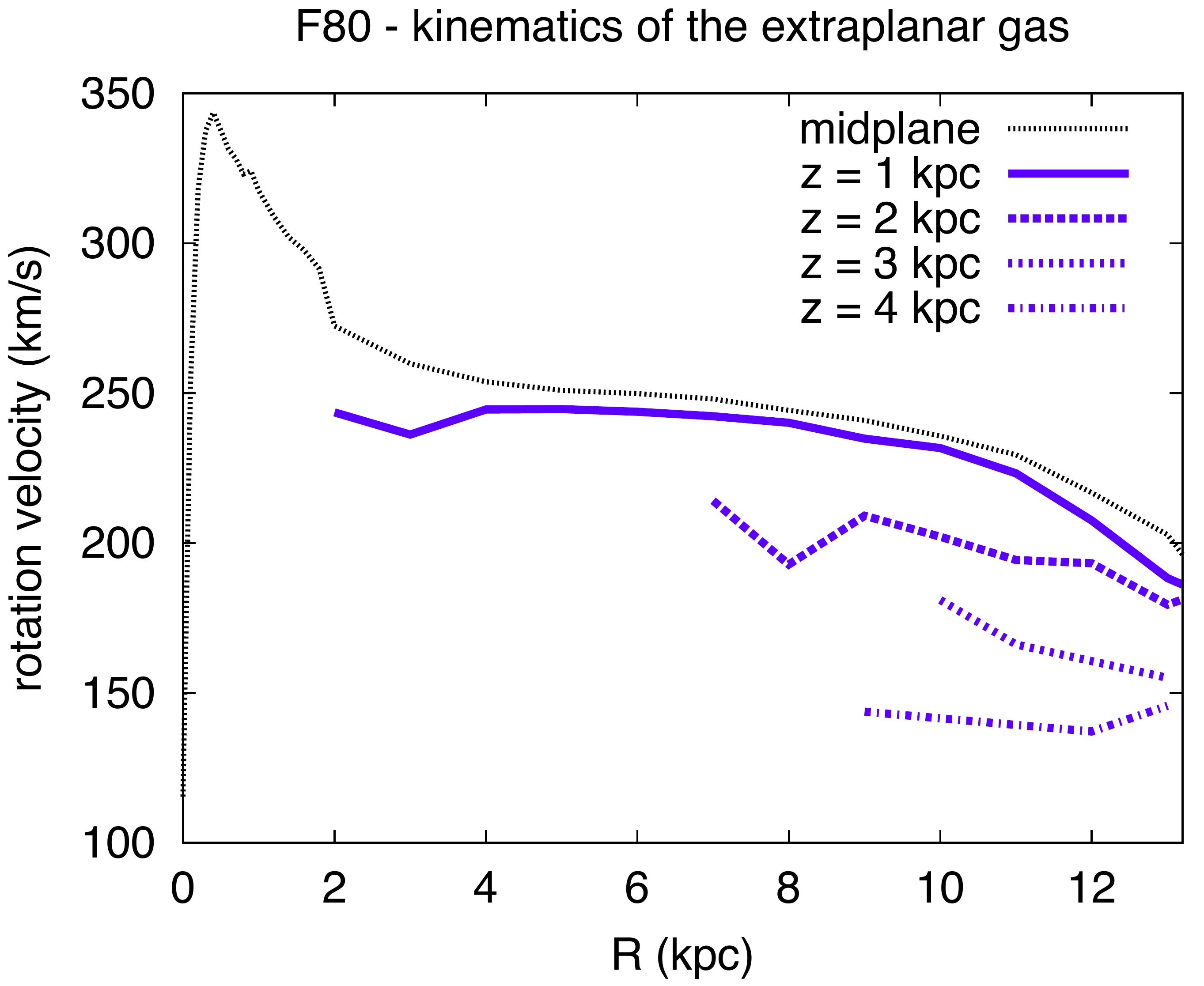}
\caption{Rotation velocity of neutral gas in model F80 at different
  heights above the midplane.}
\label{laghalo}
\end{center}
\end{figure}

F80 is the only simulated system that shows a prominent thick \hi\
disc.  We directly evaluate the rotation curves above the disc:
Fig.\,\ref{laghalo} compares the rotation curve of cold gas derived in
the midplane of F80 with those obtained at larger heights.  Rings with
less than 10 cold gas particles have been neglected.  The figure
clearly shows that the larger is the distance from the midplane, the
slower is the rotation.  At $2\kpc$ above the midplane, the difference
in rotational speed with respect to the midplane is about $30\kms$,
corresponding to a lag of $\sim15\kmskpc$, in remarkable agreement
with that found by \citet{MarascoFraternali11} for the \hi\ halo of
the Milky Way and by \citet{Oosterloo+07} for NGC 891.
We discuss the origin of this rotational lag in Section \ref{originoflag}.
Fig.\,\ref{laghalo} also shows that this component is not present in
the innermost region of the disc, pointing to an \hi\ halo with a
toroidal shape.  This is expected in the galactic fountain mechanism,
as the gravitational restoring force of the disc decreases outwards
making it easier for \hi\ gas to reach larger distances from the plane
\citep{FB06}.  We found that all the cold gas particles that form the
extra-planar gas layer have non-zero metallicity, indicating that this
material originated inside the disc.  Specifically, the metal
abundance of this component is the same as the underlying disc at all
radii, which implies that neutral gas ejected from the disc by stellar
feedback does not significantly change its galactocentric distance
during its trajectory.

In our systems, cold gas is absent at large distances from the discs
(see Fig.\,\ref{gasprofile}).  On the one hand, this is in contrast
with the results of the COS-Halo Survey \citep{Tumlinson+13}, which
revealed that cold, photo-ionized absorption systems are widespread
around galaxies \citep{Werk+13,Werk+14}.  On the other hand, the dense
\hi\ clumps that were ubiquitous in the CGM of galaxies in previous
SPH simulations \citep[e.g.][]{Kaufmann+06, Kaufmann+09} are not
present in our simulations.  This in agreement with the results of
\citet{Pisano+11}, who show the lack of massive \hi\ floating clouds
at large distances from galaxies in Local-group-like systems.

\subsection{The warm$+$hot gas} 
\label{warmhot}

By analyzing the spectra of background sources, a number of studies
have revealed the presence of absorption systems of ionized species
(\siiii, \siiv, \civ, \ovi) at $|v_{\rm LSR}|\!<\!400\kms$
\citep[e.g.][]{Wakker+03,Lehner+12}.  The global kinematics of these
absorbers is not consistent with being part of the Galaxy disc, which
makes of these systems an ionized counterpart of the classical \hi\
High-Velocity Clouds.  All these ionized species are expected to exist
at a temperature well below the Milky Way's virial temperature. Hence,
a natural explanation for these absorbers is that they are produced by
the cooling of the CGM of the Milky Way and represent a strong
evidence for accretion of ionized gas onto the Galaxy disc
\citep{LH11,Fraternali+13}.

We now compare the position-velocity distribution and the column densities of these systems
with those predicted for the warm-hot ($4.3\!<\!\log(T)\!<\!5.7$) gas
in the simulation with the highest feedback (F80) which already shows a layer of extra-planar cold gas in good agreement with the observations.
In addition, we use the full 6D information available in the simulation to determine the location of the absorbing material.
As we will remark during our analysis, our results remain valid also for the runs with lower feedback.

\subsubsection{Kinematics}
\label{abs_kinematics}

The analysis that we perform here is similar to that used by
\citet{Marasco+13} to compare their galactic fountain model with the
absorption data. In the following, we summarize the procedure adopted.
First, we use the same method described in Section \ref{extra-planar}
to produce modelcubes of the warm ($4.3\!<\!\log(T)\!<\!5.3$) and hot
($5.3\!<\!\log(T)\!<\!5.7$) gas of F80. The main difference is that
here the spatial smoothing uses the SPH kernel size of each particle
instead of a fixed angular resolution: as our data consist of a set of
absorption line measurements that are randomly distributed in the sky,
we prefer to sample the warm-hot gas in the simulated box smoothly and
extend its `filling-factor' as much as possible.  Each modelcube is
produced twice: once using only the polluted gas (i.e., with metal
abundance larger than 0), and the second time using both polluted and
unpolluted gas.  We remind the reader that, as metal diffusion is not
used and the simulation starts with no metals, particles can be
polluted only by SN explosions in the disc or stellar winds.
Therefore, the first modelcube is representative for gas that has been
affected by feedback and participates in the fountain-cycle.  A
qualitative comparison between models and data is carried out by
overlapping the position-velocity of the absorbers on top of the $l-v$
diagram of the modelcubes.  A more quantitative analysis uses an
iterative KS-test to derive the fraction of detections that is
consistent with our modelcubes \citep[for the details
see][]{Marasco+13}.

\begin{figure*}
\begin{center}
\includegraphics[width=0.48\textwidth]{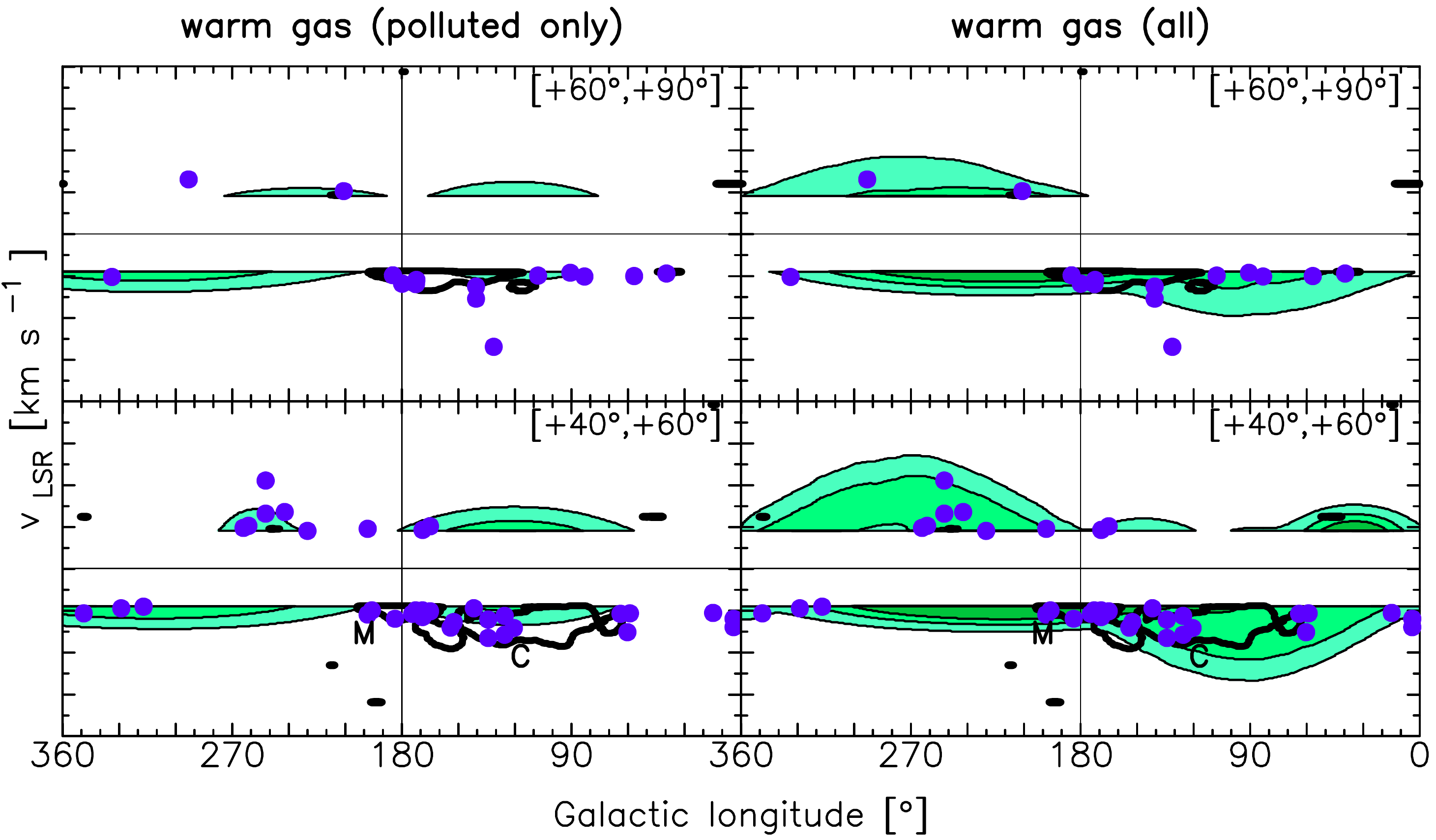}
\includegraphics[width=0.493\textwidth]{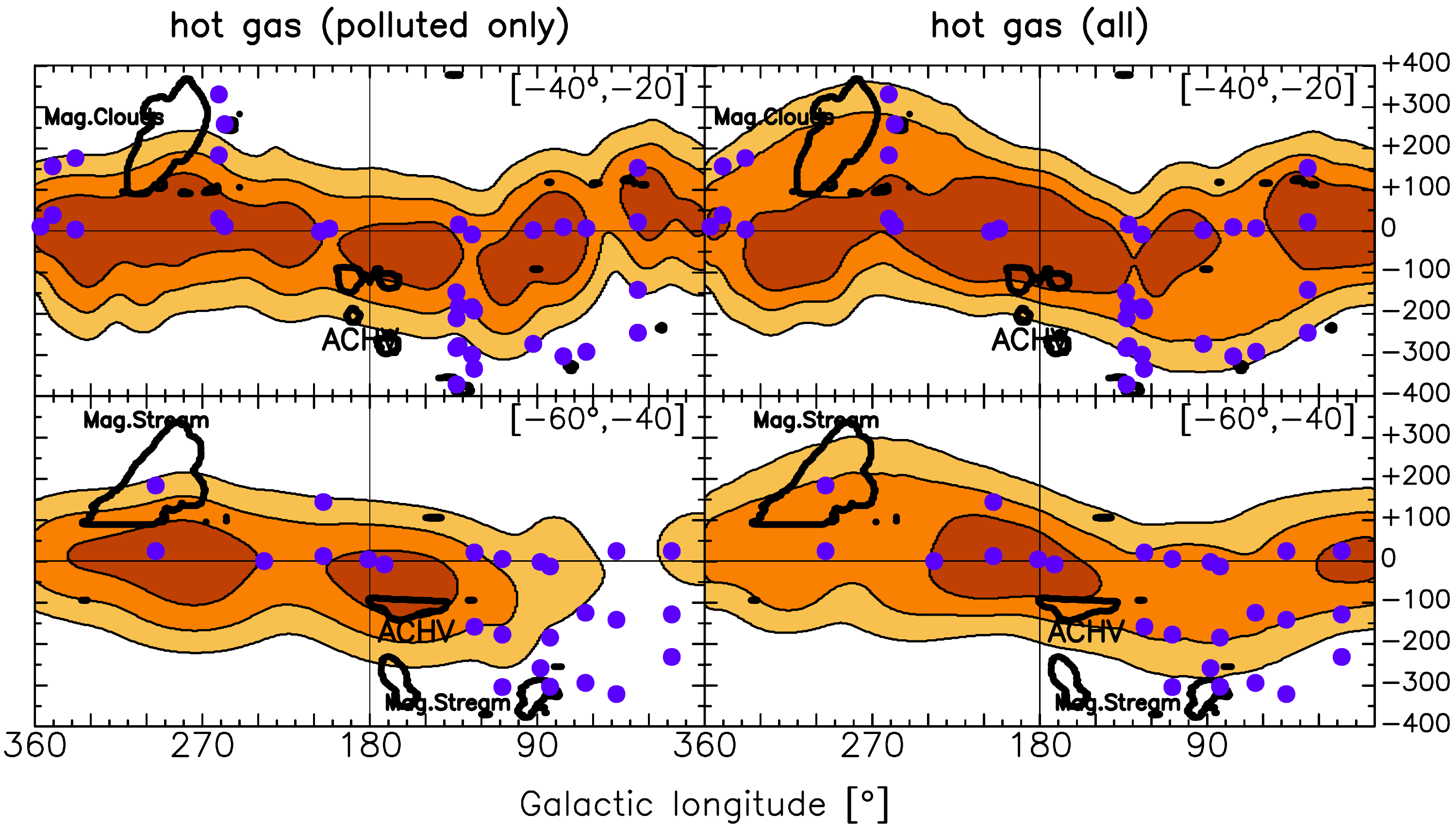}
\caption{Longitude-velocity slices in two different representative latitude bins
  (indicated on the top-right corner of each panel) for the warm
  ($4.3\!<\!log(T)\!<\!5.3$, left-hand side panels) and hot
  ($5.3\!<\!log(T)\!<\!5.7$, right-hand side panels) gas of our run F80 (filled region).
  Circles overlapped to the warm gas represents the detections from
  \citet{Lehner+12}, those overlapped to the hot gas are taken from
  \citet{Sembach+03} and \citet{Savage+03}.  Contour levels from the
  innermost to the outermost are at $68.3\%$, $95.4\%$ and $99.7\%$ of
  the total flux.  The \hi\ emission of the classical HVCs is reported
  in the various panels as the thick contours, labelled with the name
  of the respective complexes.}
\label{absorption}
\end{center}
\end{figure*}

Fig.\,\ref{absorption} shows two representative longitude-velocity
($l\!-\!v$) slices for the warm (panels on the left) and the hot
(panels on the right) gas of F80.  Slices derived for the polluted gas
alone and for the polluted+pristine material are adjacent to each
other.  The three contours shown enclose $68.3\%$, $95.4\%$ and
$99.7\%$ of the total flux of the modelcubes, by analogy with a
Gaussian distribution.  The location of the observed absorption
systems in the position-velocity diagram for these latitudes is
plotted on top of the slices.  As in \citet{Marasco+13}, the modelcube
of warm gas is compared to the absorption data of \citet{Lehner+12}
(\siiii, \siiv, \cii, \ciii, \civ) while the hot gas is compared to
the joined datasets of \citet{Sembach+03} and \citet{Savage+03}
(\ovi).  Warm material at $|v_{\rm LSR}|\!<\!90\kms$ has been
neglected as it was excluded in the observations of \citet{Lehner+12}.

The latitude bins chosen emphasize the differences between the whole
gas and the polluted material alone.  Several detections occur at
velocities that are well beyond those predicted by the warm-hot
polluted material of the simulation.  Instead, the inclusion of
pristine gas from the CGM, which rotates at a lower speed than the
disc, significantly increases the relative line-of-sight velocities
reached by the warm-hot material and produces a better overlap with
the observations.  Using our iterative KS test we find that only $6\%$
of the detections of Lehner et al. are compatible with the warm
polluted material alone, whereas this percentage increases to $45\%$
when including the unpolluted gas.  This suggests that the gas cycle
produced by stellar feedback can not be solely responsible for the
observed warm absorption features, as many absorbers have kinematics
that are more consistent with that of the CGM.  This discrepancy is
significantly mitigated when we consider the \ovi\ detections of
Sembach and Savage, as the percentage of reproduced features passes
from $31\%$ for the hot polluted material alone to $51\%$ when all the
gas is considered, indicating that a good fraction of \ovi\ around
galaxies might come from stellar feedback.  We notice that these
results do not change when metal and thermal diffusions are included
in the simulations: we derived the same $l\!-\!v$ slices of
Fig.\,\ref{absorption} using the version of F40 where diffusion is
included, and we found similar results.  Still, the models can account
at most for about half of the detections.  A possibility is that some
of the high-velocity \ovi\ absorption features observed by
\citet{Sembach+03} could originate in the CGM of nearby galaxies and
are therefore not related to the Milky Way. Following
\citet{Marasco+13}, we removed from our catalogue 41 \ovi\ features
that can be associated to external galaxies, and found that the
fraction of the remaining \ovi\ detections that is reproduced by our
model increases to $70\%$.  We therefore conclude that the kinematics
of the hot gas predicted by our runs is compatible with that shown by
the majority of the \ovi\ absorbers observed around the Galaxy.

\subsubsection{Location}
\label{abs_location}

Under the hypothesis that the gas responsible for the observed
absorption features is best represented by a mixture of warm-hot
polluted and unpolluted gas, we use the simulation to determine what
is the typical distance of the intervening absorption material.

\begin{figure}
\begin{center}
\includegraphics[width=0.5\textwidth]{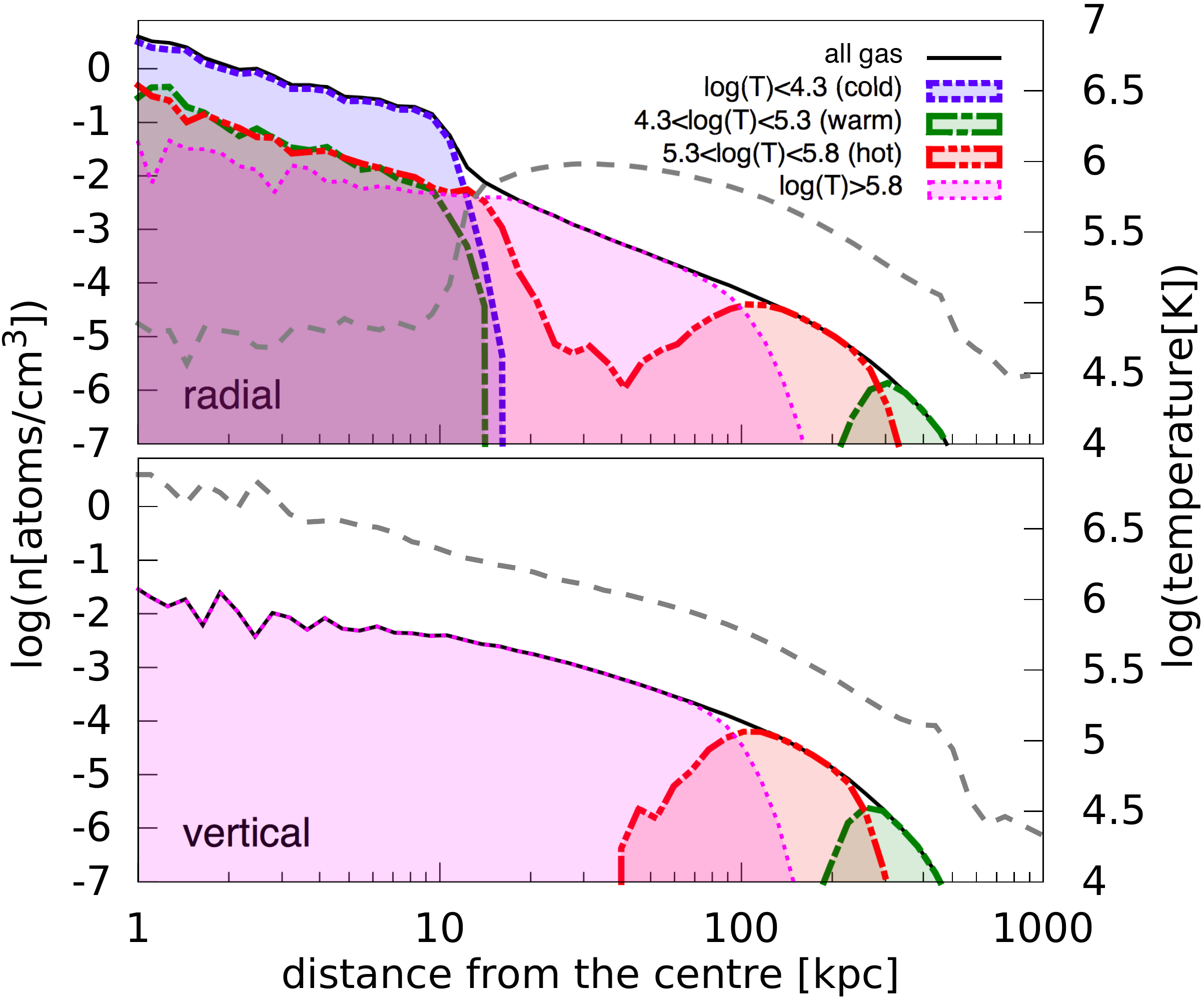}
\caption{Gas density profile (solid line, left-hand axis) and temperature profile (long-dashed  line, right-hand axis) in our model F80.
The top panel shows the profiles derived radially, i.e. intersecting the galaxy disc, while the bottom panel shows those derived vertically, i.e. along the rotation axis.
The density profiles are decomposed in four components at different temperatures, as indicated in the top panel.}
\label{gasprofile}
\end{center}
\end{figure}

Fig.\,\ref{gasprofile} shows both the radial (i.e., parallel to the
disc) and the vertical (perpendicular to the disc) gas volume density
profiles in model F80.  The former has been derived by considering all
gas particles in a cone centred on the system centre with an aperture
of $5\de$ above and below the midplane, which allows us to intersect
the region of the disc and exclude any contamination from gas above
it, while for the latter we used a cone of $30\de$ around the rotation
axis of the system.  We used different lines and colours to represent
the various phases of the gas.  Cold gas is present only inside the
galaxy disc, as it appears only in the inner $\sim10\kpc$ of the
radial profile.  The vertical profile does not show the presence of
cold material even in the innermost regions, where cold extra-planar
gas is located.  This happens because, as discussed in Section
\ref{extra-planar}, this layer has a toroidal shape and therefore it is
missed by the cone of view that we adopt.  Warm and hot gas are
instead located both in the disc and in the CGM.

As the kinematics of the warm gas coming from the disc is not in
agreement with the observations (Section \ref{abs_kinematics}) we must
focus on the warm material in the CGM, which appears to be located at
distances between $150\kpc$ and $500\kpc$ from the centre, beyond the
hot gas.  As we are adopting a simple temperature threshold to
discriminate between warm and hot material, this can be interpreted in
terms of a temperature gradient in the CGM.  The temperature of the
CGM in the inner region of the system is larger than that in the
outskirts (long-dashed line in the bottom panel of Fig.\,\ref{gasprofile}),
inducing a spatial separation between warm and hot material and
creating a deficit of warm gas between $20$ and $200\kpc$.  However,
gas at this temperature is commonly seen in absorption within
$150\kpc$ from galaxies \citep[e.g.][]{Tumlinson+13}.  Furthermore, in
the Milky Way, there is evidence that most of the warm absorbers are
located in the lower halo, within a few kpc from the Galaxy disc
\citep{LH11,Wakker+12,Marasco+13}.

The hot gas in F80 extends radially from the system centre up to more
than $\sim300\kpc$ from the disc, although it is absent in the inner
$40\kpc$ in the direction perpendicular to the midplane.  This is in
agreement with the \ovi\ impact-parameters derived in external
galaxies: both \citet{Stocke+06} and \citet{WakkerSavage09} found that
low-redshift intergalactic \ovi\ originates within 500 kpc from bright
galaxies.  In order to verify whether the absorption systems are
widespread in the CGM or are confined within a shell at a given
distance from the system centre, we proceed as follows: we consider
the angular position and the $v_{\rm LSR}$ of the \ovi\ absorbers
found by \citet{Sembach+03} and \citet{Savage+03}, and we determine at
what distance from the disc the hot gas particles of F80 with the same
position and $v_{\rm LSR}$ are located.  We found that there is no
preferential location for these particles, $\sim70\%$ of them lie
between 20 and 200 kpc from the centre.

Our findings suggest that the location of the hot gas predicted by
model F80 is in agreement with the absorption-line observations, while
that of the material at a lower temperature is not.  We stress that
our results are valid also for the models with lower feedback, as
their density and temperature profiles are very similar to those of
F80.  The only exceptions are found in the inner region of the radial
profile where, as the feedback decreases, the hot material inside the
disc gets increasingly replaced by warm gas in response to the lower
amount of energy injected into the ISM.  One may wonder whether our
findings are affected by the lack of metal and thermal diffusion, as
the mixing between the different gas phases can alter the temperature
profile of the CGM.  To address this question, we compared the gas
temperature and density distribution in F40 with those derived by
re-simulating the same system with thermal and metal diffusion
included.  We found no differences, except that the low-density hot
gas located between $20$ and $80\kpc$ from the center disappears,
likely because it gets thermalized by the (hotter and denser)
surrounding coronal gas.  It is possible that warm or cold material
arises in the inner hundreds of kpc from the discs only in a full
cosmological context as a consequence of late-time filamentary
cold-mode accretion, as suggested by high-resolution cosmological
simulations of Milky Way-like systems
\citep{KeresHernquist09,Joung+12, Stinson+12}.  Insofar as our
findings point towards this scenario, we remind the reader that the
evolution of the CGM in the systems depends in principle on the
initial conditions of the simulation.  Adopting an isothermal corona
\citep[e.g.][]{Binney+09}, a lower baryonic fraction \citep[motivated
by X-ray observations, see][]{AndersonBregman10}, a different
distribution of the initial angular momentum, the presence of
substructure, or a different feedback recipe are all choices that
might have lead to different results.

\subsubsection{Column density}

Since the hot phase of the gas in the simulation is the one that best
matches the position and the kinematics of the observed \ovi\
absorption systems, it is interesting to test whether the amount of
hot gas in the simulations is consistent with the observed \ovi\
column densities.  The comparison between the simulations and the
observations is not straightforward, as the former are not designed to
keep track of the various heavy elements nor to determine the gas
ionization balance.  In the following, we attempt to determine the
\ovi\ column density in a generic line of sight $N_{\rm O_{\rm
    VI}}(l,b)$ by using the information that is available for each
particle, namely its total mass, temperature and metal abundance.

In the models, the \ovi\ mass $m_{\rm O_{\rm VI}}$ associated to each gas
particle of total mass $m$ can be written as
\begin{equation}\label{ovimass}
%  m_{\ovicap} = m \left(\frac{m_{\ovicap}}{m_{\rm O}}\right)\left(\frac{m_{heavy}}{m}\right)\left(\frac{m_{\rm O}}{m_{\rm H}}\right)\left(\frac{m_{heavy}}{m_{\rm H}}\right)^{-1}
m_{\rm O_{\rm VI}} = m \left(\frac{m_{\rm O_{\rm VI}}}{m_{\rm O}}\right)\left(\frac{m_{heavy}}{m}\right)\left(\frac{m_{\rm O}}{m_{heavy}}\right)
\end{equation}
where $m_{\rm O}$ is the oxygen mass associated to the particle and
$m_{heavy}$ is the mass of the elements heavier than helium.  The
three bracketed terms in eq.(\ref{ovimass}) can be determined
separately.  We assume collisional ionization equilibrium (CIE) to
evaluate $(m_{\rm O_{\rm VI}}/m_{\rm O})$ as a function of the
temperature of the particle, by using the oxygen ionization-balance
table of \citet{SutherlandDopita93}.  Note that the assumption of CIE
should be solid for \ovi\ systems in the halo of galaxies, as
discussed by \citet{Sembach+03}.  The quantity $(m_{heavy}/m)$ is the
metal abundance of the particle, which is known in all our
simulations.  Finally, we assume solar abundance ratios for our gas
and set $(m_{\rm O}/m_{heavy})\!=\!(m_{\rm O}/m_{\rm
  H})_{\odot}\times(m_{heavy}/m_{\rm
  H})_{\odot}^{-1}\!=\!10^{-3.27}\times(0.0191)^{-1}$
\citep{Lodders10}.

We use a procedure similar to that described in Section \ref{HIdisc}
to build a modelcube where each voxel $I(l,b,v)$ contains information
on the \ovi\ column density per unit velocity at the position $(l,b)$
of the simulated sky.  The \ovi\ column density at position $(l,b)$
will be simply given by
\begin{equation}
\label{ovicd}
N_{\rm O_{\rm VI}}(l,b) = \int_{v_1}^{v_2}I(l,b,v)\,{\rm d}v
\end{equation}
where $v_1$ and $v_2$ are two generic line-of-sight velocities.
Following \citet{Marasco+13}, we use eq.\,(\ref{ovicd}) to compute
$N_{\rm O_{\rm VI}}$ in all line-of-sights $(l,b)$ where \ovi\
detections have been found in the combined datasets of
\citet{Sembach+03} and \citet{Savage+03}.  The integral in
eq.\,(\ref{ovicd}) is calculated between $v-\frac{2}{\sqrt2}b_w$ and
$v+\frac{2}{\sqrt2}b_w$, where $v$ is the mean velocity and $b_w$ is
the line-width of the observed \ovi\ absorption feature.  By
considering all the \ovi\ column densities computed in the various
line-of-sights, we find for model F80 an average value of
$\avg{\log{N_{\rm O_{\rm VI}}}}\!=\!13.7\pm1.03$, compatible with the
value derived from the observations, $14.16\pm0.34$.  This result
changes little when metal and thermal diffusions are included, as the
average \ovi\ column density in the version of F40 that includes
diffusion is $14.48\pm1.45$.  Considering all the uncertainties, we
conclude that the systems show a layer of hot gas whose kinematics,
location and column densities are compatible with what is inferred for
the Milky Way and external galaxies from \ovi\ measurements.

\section{Discussion}
\label{discussion}

\subsection{The impact of stellar feedback}\label{feedbackimpact}
\begin{figure}
\begin{center}
\includegraphics[width=0.45\textwidth]{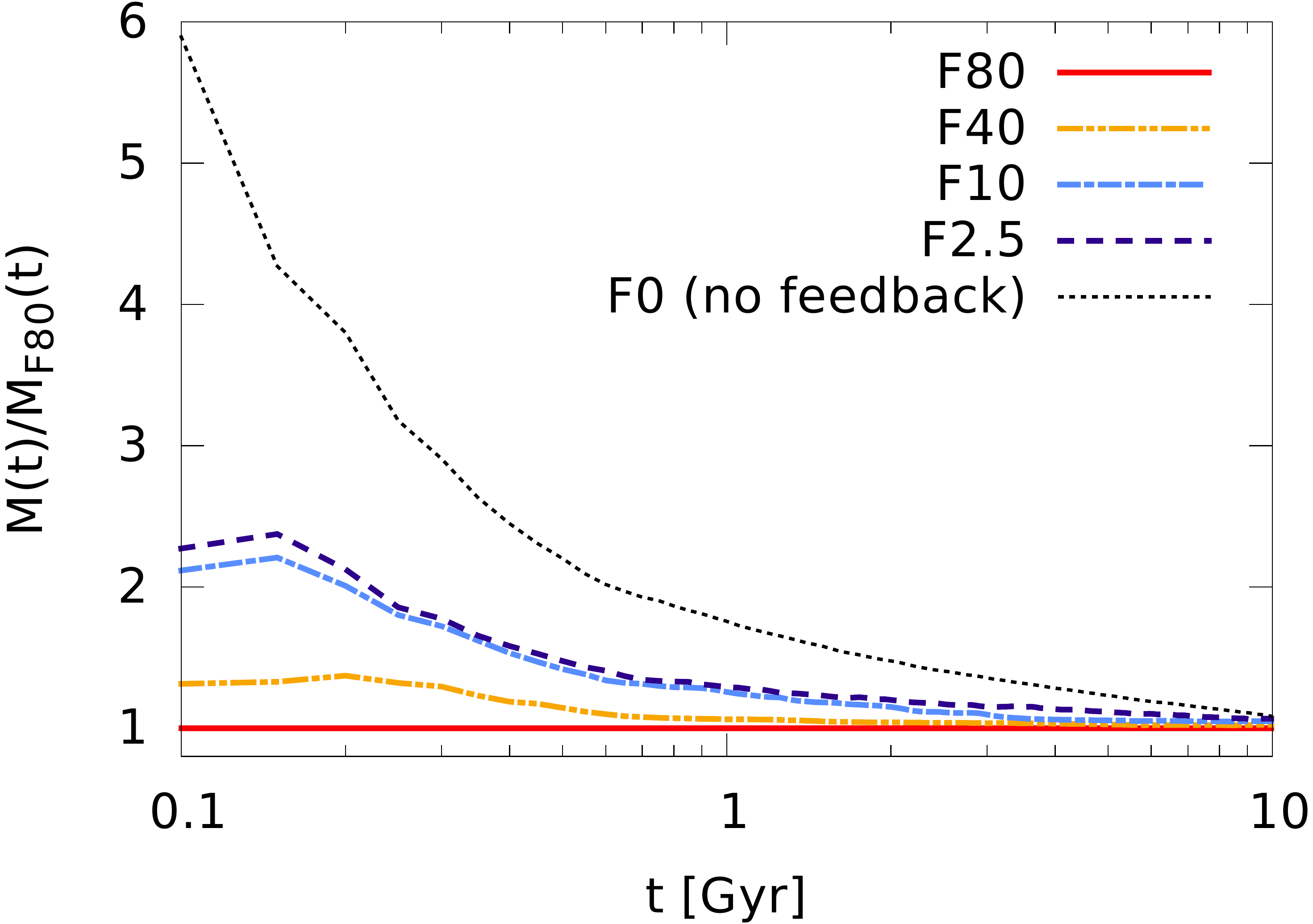}
\caption{Total stellar mass assembled in our runs, normalized to the
  stellar mass assembled by F80, as a function of time.  The x-axis is
  in logarithmic scale to emphasize the early stage of the evolution.}
\label{cumulative}
\end{center}
\end{figure}

One of the most puzzling features of the simulated galaxies is that,
as shown by Fig.\,\ref{SFH}, all star formation histories are very
similar despite the fact that $E_{\rm SN}$ changes significantly from
simulation to simulation.  This result dramatically differs from that
found by S06, i.e. that the injection of energy into
the ISM slows down the star formation processes and therefore the same
amount of fuel gets smoothly consumed on longer timescales.  The main
difference between our work and that of S06 is that the
systems have coronae and hot-mode accretion happens all the time.  Our
findings indicate that, at least in the range of feedback energies
studied in this work, gas accretes onto the discs - and star formation
proceeds - in a similar fashion for all the simulated systems.  But is
stellar feedback affecting the hot-mode accretion at all?  To answer
this question, we ran another simulation (F0) adopting the same
initial conditions as the other simulations, but with stellar feedback
switched off altogether.  Fig.\,\ref{cumulative} shows the total
stellar mass assembled in all our runs (including F0), normalized to
the stellar mass assembled by F80, as a function of time.  Clearly,
the lower the feedback is, the quicker the stellar mass is assembled.
In particular, throughout the first Gyr, F0 builds a stellar component
that is several times more massive than that of F80 (consistent with
the results of S06), while the difference between the
runs with intermediate feedback and F80 is less pronounced.  This
discrepancy slowly fades away with time: after 10 Gyr of evolution the
total stellar mass assembled is virtually the same in all systems.
Our interpretation is that, at the beginning of the simulation,
stellar feedback quickly removes a significant fraction of gas that
was piling up at the centre of the system, quenching star formation in
the first few hundreds of Myr.  A key point is that, in the feedback
energy range explored in the models, this material remains
gravitationally bound to the system, thus it slowly re-accretes onto
the disc and takes part in the process of star formation. After 10
Gyr, the total amount of stars formed in the simulations with feedback
and in that without feedback is similar, as the total amount of fuel
available does not differ.  As F0 has used most of its star forming
fuel at the beginning of its evolution, the amount of neutral hydrogen
left at $t\!=\!10\Gyr$ is $3.8\times10^{8}\msun$, $\sim13\%$ of that
available in the disc of F80 at the same time.

It must be pointed out that the parameter $E_{\rm SN}$, on which our analysis is focussed, is only one of the many ingredients that determine the global `effectiveness' of stellar feedback in simulations. 
For instance, \citet{Governato+10} realized the importance of the density threshold of gas eligible for being star-forming in simulations of dwarf galaxies: the larger the threshold, the more efficient the feedback is in preventing the formation of a prominent stellar bulge by producing strong outflows from the center that remove low angular momentum material.
To explore this scenario, we re-ran our simulation F40 by increasing the star formation density threshold from our `fiducial' value of $0.1$ cm$^{-3}$ (S06) to $100$ cm$^{-3}$.
We found that the final ($t=10$ Gyr) stellar mass of this new system is similar to that reported in Table \ref{snapshots}, but the central stellar surface density is about $40\%$ smaller than that of our fiducial run F40, indicating that now feedback is slightly more efficient in preventing a mass stockpiling at the system centre. 
The total \hi\ mass, also, is slightly larger ($3.3\times10^9\msun$).
The change in kinematics is however negligible, as the peak rotational velocity of this new system is only $10\kms$ lower than that of our fiducial run.
A detailed analysis of the impact of the density threshold on our systems is beyond the purpose of our study, but these results may indicate that massive systems like those studied in this paper are relatively less affected by large variation of this quantity. 
%This experiment confirms a well known issue, i.e. that the detail of a feedback recipe can have a large impact on the outcome of a simulation \citet[e.g.][]{Scannapieco+12}.

Finally we would like to stress that, even though our runs adopt a relatively high $E_{\rm SN}$, in state-of-the art cosmological simulations even larger values are required or additional feedback mechanisms (like radiative feedback from young stars or AGN feedback) are included in order to match the observations \citep{Hopkins+14,Schaye+15}.
Our system with highest feedback shows a layer of extra-planar \hi\ which is less massive and extended than that observed in NGC 891, a galaxy with a similar star formation rate and comparable total \hi\ and stellar masses \citep{Oosterloo+07, Fraternali+11}.
In future studies, it will be interesting to explore values of $E_{\rm SN}$ larger than those considered in this work.

\subsection{The origin of the rotational lag}\label{originoflag}
\begin{figure}
\begin{center}
\includegraphics[width=0.49\textwidth]{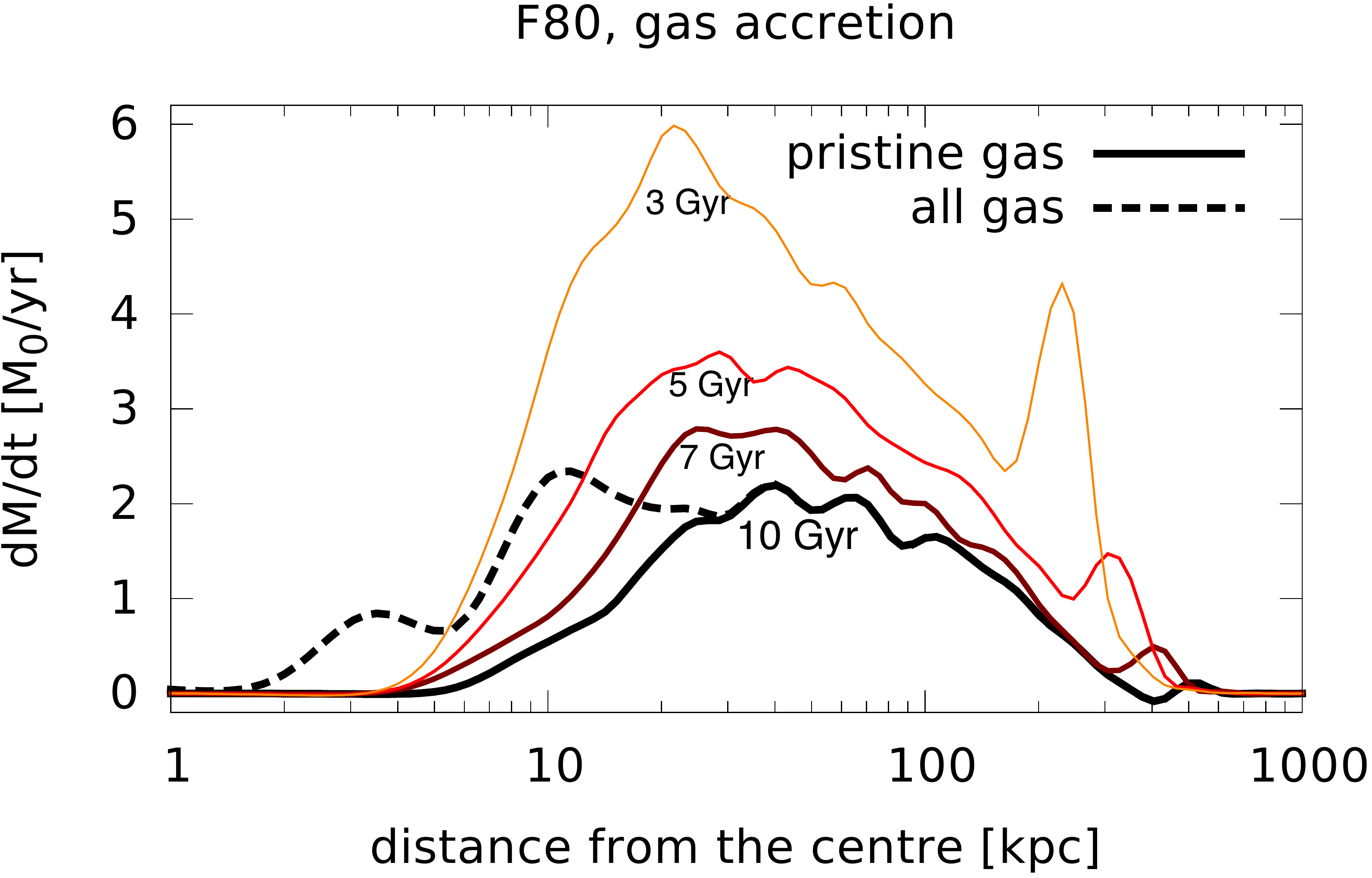}
\caption{Net gas flow (defined in eq.\,\ref{accretion}) as a function
  of the distance from the system centre derived at four different times in our model F80. Positive values indicate accretion. The solid lines represent pristine gas only, the dashed line (shown only at $t\!=\!10\Gyr$) includes polluted material as well.}
\label{gasflow}
\end{center}
\end{figure}

In Sections \ref{morphology} and \ref{extra-planar} we discussed the
morphology and the kinematics of the extra-planar cold gas of the model
with highest feedback, F80. To our knowledge, this is the first time
that a global hydrodynamical simulation of a disc galaxy shows the
presence of a thick \hi\ disc produced by stellar feedback with
kinematics in agreement with the observations.
  \citet{Melioli+09} used adaptive mesh refinement hydrodynamical simulations to follow the
dynamical evolution of a galactic fountain flow powered by $\sim100$
OB associations in a Milky Way-like galaxy.  Both the supernova rate
and energy input used in their work are comparable to those adopted in
model F80.  \citet{Melioli+09} found that gas from the ISM gets lifted
up to a few kpc above the disc and falls back approximately to the
same radius, in agreement with our results.  However, they could not
reproduce the lag in rotational velocity that is observed in the \hi\
halo of real galaxies, and suggested that the galactic fountain flow
interacts with extragalactic material inflowing onto the galaxy with low angular momentum to
produce these lagging kinematics. 
The same conclusions were
previously found by \citet{FB08} (hereafter FB08) using a dynamical model of the
galactic fountain. 
In our simulations, a natural source of gas inflow
is the gas corona. Following \citet{Peek+08}, the net
(inflow-outflow) gas accretion in a spherical shell at a distance $r$
from the system centre can be determined in the models as
\begin{equation} \label{accretion}
\frac{{\rm d}M(r)}{{\rm d}t} = \sum_{i=0}^{n(r)} \frac{m_i\,v_{r,i}}{\Delta r}
\end{equation}
where $m_i$ is the mass of the $i$-th gas particle, $v_{r,i}$ is the
radial (i.e. directed towards the centre) component of its velocity,
$\Delta r$ is the thickness of the shell and the sum is extended to
all $n(r)$ particles inside that shell.  Fig.\,\ref{gasflow} shows the
net (inflow-outflow) gas flow as a function of the distance from the
system centre in F80, derived at four different times by evaluating eq.(\ref{accretion}) in a
series of shells, each one located at a distance $10^{0.03k}\kpc$
($0\!\le\!k\!\le\!100$) from the centre. At $t\!=\!10\Gyr$, pristine gas (solid
line) between $30$ and $300\kpc$ is globally inflowing at a rate of
about $2\msunyr$.  At smaller distances, the inflow of pristine
material appears to vanish as particles get contaminated by supernova
feedback from the disc.  However, when the total (polluted + pristine)
gas is considered, the inflow appears to proceed down to the system
centre (dashed line). In addition to the gas accreted from the
corona, the discs of our systems are re-fuelled also by stellar winds,
which return to the ISM about $25\%$ of the gas converted into stars.
Therefore, we estimate that the rate at which new gas is reverted to
the disc of F80 at $t\!=\!10\Gyr$ is about $3\msunyr$, which comes
relatively close to the SFR of this system ($4.2\msunyr$).  We also
point out that in all the models (including F0) the mass flow profile is very
similar, suggesting that stellar feedback does not have a big impact
on the way the models are accreting gas from the CGM.
Fig.\,\ref{gasflow} shows also that the gas accretion rate was higher at early times, peaking around $3.5\msunyr$ at $t=5\Gyr$ and $6\msunyr$ at $t=3\Gyr$.
Interestingly, this is in good agreement with the inflow rates derived from simulations of Milky Way-like galaxies in a full cosmological context \citep[e.g.][]{Brook+14}.

\begin{figure}
\begin{center}
\includegraphics[width=0.45\textwidth]{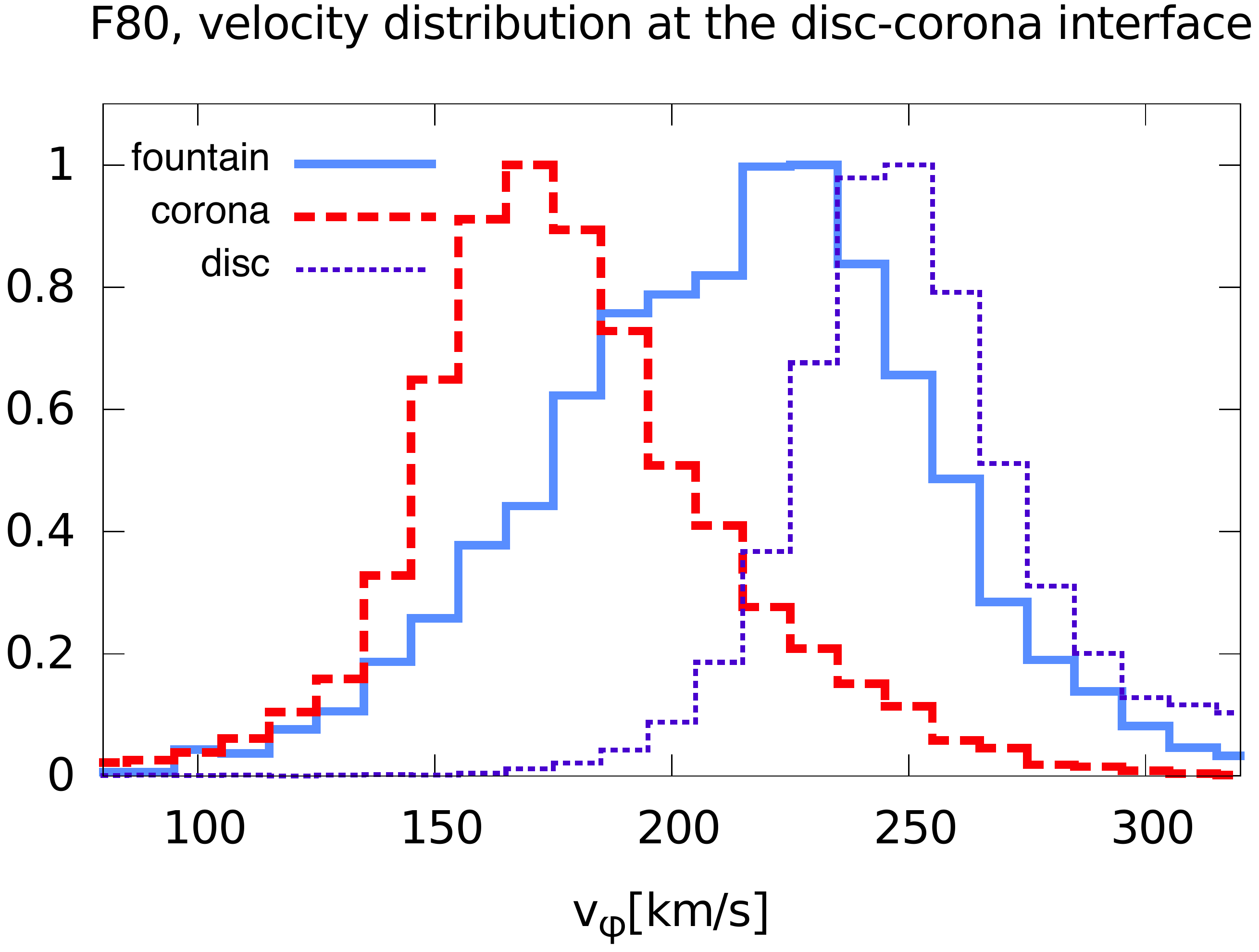}
\caption{Azimuthal velocity distribution of gas particles located at the disc-corona interface ($1\!<\!|z|\!<\!4\kpc$) in our run F80, at $t\!=\!10\Gyr$. The solid line shows the material ejected from the disc by stellar feedback, defined as having $\log(Z/Z_{\odot})\!>\!-1.2$ (see text), while the dashed line shows the coronal gas. For comparison, the dotted line shows the velocity distribution of cold gas in the disc. Each distribution is normalised to its peak value.}
\label{vphiinterface}
\end{center}
\end{figure}

In model F80, it is not straightforward to determine the details of
the interaction between the cold, metal-rich gas lifted from the disc
and the hot coronal material inflowing to the centre, as pristine gas
particles are absent in the region where extra-planar cold gas exists.
This is likely caused by stellar winds from evolved stars in the thick stellar disc, which inject mass and metals into the surrounding gas particles (S06). In fact, pristine gas in F80 appears only above $4\kpc$ from the disc, precisely where the stellar component fades out.
This implies that, at the disc-corona interface, we can not anymore use the simple polluted/unpolluted criterium to determine to which component a gas particle belongs.  
However, we found that the metallicity distribution of gas within a few kpc from the disc is bimodal, with two clear peaks located around $Z\!=\!Z_{\odot}$ and $Z\!=\!0.01Z_{\odot}$ and a separation occurring roughly at $\log(Z/Z_{\odot})\!=\!-1.2$.
For the purpose of investigating the disc-corona interplay, we interpret the gas at $\log(Z/Z_{\odot})\!>\!-1.2$ as material ejected from the disc by stellar feedback and taking part in the galactic fountain gas cycle, and the gas at $\log(Z/Z_{\odot})\!<\!-1.2$ as coronal gas which has never been part of the disc but is polluted by thick disc stars.
Our interpretation is corroborated by the fact that gas particles at $\log(Z/Z_{\odot})\!<\!-1.2$ are preferentially located outside the disc, while those at a higher metallicity are mainly located inside it.
Fig.\,\ref{vphiinterface} shows the rotation velocity distribution for these two components in the region where extra-planar cold gas is present, $1<|z|<4\kpc$, and compares it with that of the underlying \hi\ disc ($|z|\!<\!0.5\kpc$).
Clearly, coronal gas rotates much slower than the material in the disc, whereas the galactic fountain gas distributes in between these two components.
The kinematics of the coronal gas is not surprising since, given the high temperature, this material is partially pressure-supported and therefore must rotate at a velocity lower than that of the circular orbit.
Thus, a very likely interpretation is that metal-rich gas expelled from the disc interacts hydrodynamically with the slow-rotating and inflowing corona, loses angular momentum and produces a rotation velocity gradient in agreement with the \hi\ observations.
This mechanism is qualitatively similar to that proposed by FB08.

We stress that the \emph{loss} of angular momentum experienced by fountain gas is likely to happen only at the disc-corona interface, where the specific angular momentum of the coronal gas is less than that of the disc material.
However, powerful feedback events can eject gas up to distances of several tens or hundreds of kpc, where this condition is reversed: this material can \emph{gain} some angular momentum from the corona before being re-accreted back to the disc \citep{Brook+12,Ubler+14}. 
It is possible that such a mechanism occurs also in our runs since polluted gas is present up to $60\kpc$ from the system centres.

The effects of the disc-corona interplay on the accretion of coronal gas have been studied in a series of papers \citep{Marinacci+11, MFB12, Fraternali+13}. 
In these works, a combination of dynamical models and high-resolution hydrodynamical simulations have been used to propose a scenario where the cooling of the lower corona is enhanced by stellar feedback from the disc via the galactic fountain mechanism. 
This process, known as `supernova-driven gas accretion', can hardly take place in our simulations
as the resolution is not adequate to describe the turbulent mixing
between these two gas layers.

\section{Conclusions}
\label{conclusions}

We analysed the outcome of four N-body+SPH simulations of a
Milky-Way like system produced by the radiative cooling of a rotating
gaseous corona embedded in a dark matter halo.  Each simulation uses
the same initial conditions, but differs by the amount of energy per
supernova, $E_{\rm SN}$, injected into the interstellar medium, which
spans a factor of 32 from the lowest feedback model
($2.5\times10^{49}$ erg/SN) to the highest feedback model
($80\times10^{49}$ erg/SN).  We studied how the morphology and the
kinematics of these systems after 10 Gyr of evolution compare to those
of the Milky Way.  In addition, we performed a direct comparison with
the \hi\ and absorption-line data of the Milky Way by simulating an
observation of the simulated systems from within the galaxy.  Our main
results are the following:
\begin{itemize}

\item in the range of input energies considered, stellar feedback has
  almost no impact on the star formation histories of the model
  galaxies;

\item stellar feedback significantly affects the vertical distribution
  of cold gas, as high feedback produces thicker discs with larger
  velocity dispersion;

\item the simulation with the highest feedback (F80) shows a prominent
  layer of extra-planar cold gas. The kinematics of this medium is
  consistent with that derived in real galaxies like the Milky Way and
  NGC 891, although its column density is about one order of magnitude
  lower. This finding supports the idea that such features are
  produced by feedback activity from star-forming discs.

\item The location, kinematics and typical column density of the hot
  ($5.3\!<\!\log(T)\!<\!5.8$) gas in F80 are consistent with those
  inferred in the Milky Way and external galaxies via \ovi\ absorption
  line studies. Gas at lower temperatures is commonly observed in
  absorption around galaxies, but it is instead scarce in the CGM of our simulated systems.

\end{itemize}
In addition, we found that the increase of $E_{\rm SN}$ affects the
stellar component in two ways: a) it systematically enhances the
central stellar surface density, and consequently the concentration of
dark matter in the galaxy centres; b) it reduces the thickening and
the velocity dispersion of the stellar disc.

\section*{Acknowledgments}

The authors thank the anonymous referee for insightful comments.
AM and TvdH acknowledge support from the European Research Council under the
European Union's Seventh Framework Programme (FP/2007-2013) / ERC
Grant Agreement nr. 291531.  VPD was supported by STFC Consolidated
grant \#~ST/J001341/1.  FF acknowledges financial support from PRIN
MIUR 2010-2011, project ``The Chemical and Dynamical Evolution of the
Milky Way and Local Group Galaxies'', prot. 2010LY5N2T.  Simulations
in this paper were carried out using the computational facility at the
University of Malta procured through the European Regional Development
Fund, Project
ERDF-080\footnote{http://www.um.edu.mt/research/scienceeng/erdf\_080)}.

\bibliographystyle{mn2e}
\bibliography{feedback}{}
\bsp

\label{lastpage}
\end{document}